\documentclass[aps,prd,showpacs,superscriptaddress,preprintnumbers]{revtex4-2}
\usepackage{graphicx,float,wrapfig,subfigure}
\usepackage{amsfonts,amsmath,amssymb,amstext}
\usepackage{latexsym}
\usepackage{bm}
\usepackage{color}
\usepackage[normalem]{ulem}
\usepackage{MnSymbol}
\usepackage[colorlinks=true,linktocpage=true,linkcolor=blue,citecolor=blue]{hyperref}
\usepackage[sep=5pt, offset=1.5em]{simpler-wick}
\usepackage{orcidlink} 
\usepackage{slashed}

\newcommand{\be}{\begin{equation}}
\newcommand{\ee}{\end{equation}}
\newcommand{\ba}{\begin{eqnarray}}
\newcommand{\ea}{\end{eqnarray}}

\definecolor{red}{rgb}{0.7,0,0}
\definecolor{green}{rgb}{0,0.5,0}

\begin{document}
  
\title{Magnetoviscosity of relativistic plasma}
 
 \author{Ritesh Ghosh}
 \email{riteshghosh283@gmail.com}
 \affiliation{College of Integrative Sciences and Arts, Arizona State University, Mesa, Arizona 85212, USA}
\affiliation{Institute of Physics, Academia Sinica, Taipei 11529, Taiwan}

 \author{Igor A. Shovkovy}
 \email{igor.shovkovy@asu.edu}
 \affiliation{College of Integrative Sciences and Arts, Arizona State University, Mesa, Arizona 85212, USA}
 \affiliation{Department of Physics, Arizona State University, Tempe, Arizona 85287, USA}
 
 \date{\today}
 
 \begin{abstract}
Using first-principles quantum field-theoretical methods, we investigate the shear and bulk viscosities of strongly magnetized relativistic plasmas. The analysis is performed within the weak-coupling approximation and utilizes known results for the fermion damping rates in the Landau-level representation, $\Gamma_{n}(p_{z})$, which are dominated by one-to-two and two-to-one processes in the presence of a strong magnetic field. The transverse and longitudinal components of the viscosities are derived using Kubo's linear response theory. Our results reveal a pronounced anisotropy in both shear and bulk viscosities induced by the magnetic field. In the case of an electron-positron plasma, where the weak-coupling approximation is well justified, the dimensionless longitudinal shear viscosity $\eta_{\parallel}/T^3$ increases rapidly with the magnetic field strength, while the transverse component $\eta_{\perp}/T^3$ decreases and can even drop below the KSS bound at sufficiently large fields. In contrast, both the dimensionless longitudinal and transverse bulk viscosities, $\zeta_{\perp}/T^3$ and $\zeta_{\parallel}/T^3$, initially rise from small values, reach a maximum, and then gradually decrease toward zero. We find that the bulk viscosity is highly sensitive to the longitudinal and transverse components of the sound velocity, particularly at high magnetic fields, indicating that its quantitative values should be interpreted with caution. We also calculate an additional cross viscosity, which is negative and whose magnitude increases with the magnetic field strength. Finally, we discuss the physical implications of these magnetoviscosity results in the contexts of magnetar physics and the strongly magnetized quark-gluon plasma produced in heavy-ion collisions.
\end{abstract} 
 
 \maketitle

\section{Introduction}
\label{Introduction}

In many extreme environments, from the primordial plasma of the early Universe to the quark-gluon plasma created in heavy-ion collisions, matter exists in the form of a relativistic plasma, where particles move at average velocities approaching the speed of light. Such plasmas often coexist with strong magnetic fields that play a crucial role in profoundly altering their physical behavior \cite{Grasso:2000wj,Kaspi:2017fwg,Skokov:2009qp,Voronyuk:2011jd,Bloczynski:2012en}. These fields influence a wide range of properties, including thermodynamics, transport coefficients, collective excitations, emission characteristics, and the development of instabilities. Consequently, they leave observable imprints on relativistic matter in cosmological, astrophysical, and laboratory contexts \cite{Durrer:2013pga,Vachaspati:2020blt,Turolla:2015mwa,Mereghetti:2015asa,Gorbar:2021tnw,Kharzeev:2024zzm}.

One specific example is the electron-positron plasma, commonly found in the magnetospheres of magnetars, which are neutron stars with magnetic fields exceeding $10^{14}~\mbox{G}$ \cite{Turolla:2015mwa,Kaspi:2017fwg}. In these regions, the strong magnetic fields not only coexist with but also facilitate the formation of electron-positron pairs in vacuum gaps, creating a highly relativistic magnetized plasma \cite{Goldreich:1969sb,Ruderman:1975ju}. Another extreme example is the quark-gluon plasma (QGP), which forms at temperatures on the order of hundreds of megaelectronvolts ($\sim 10^{12}~\mbox{K}$), where quarks and gluons become deconfined from hadrons \cite{Yagi:2005yb}. This state of matter existed briefly after the Big Bang \cite{Grasso:2000wj}, and tiny droplets of QGP are now recreated in relativistic heavy-ion collisions \cite{STAR:2005gfr,PHENIX:2004vcz,PHOBOS:2004zne}. Moreover, non-central collisions generate extremely strong, short-lived magnetic fields \cite{Skokov:2009qp,Voronyuk:2011jd,Bloczynski:2012en}, which can profoundly modify the dynamics and observable features of the QGP.

An essential aspect of understanding physical properties of relativistic plasmas is to study their transport properties, which describe how the system responds to external perturbations or internal gradients. Among other characteristics, shear and bulk viscosity play critical roles. The shear viscosity characterizes the diffusion of momentum transverse to a velocity gradient, thereby governing how efficiently the plasma can redistribute momentum and maintain flow. The bulk viscosity, on the other hand, quantifies the system's response to uniform compression or expansion, reflecting how the internal degrees of freedom equilibrate during volume changes. Conceptually, the viscosity determines how ``perfect" the fluid is. 

In the context of heavy-ion physics, the observation of collective hydrodynamic flow provides strong evidence that the QGP behaves as an almost ideal fluid with very low viscosity \cite{Heinz:2013th}. This behavior is consistent with the picture of the QGP as a strongly coupled medium. Although most theoretical studies support this conclusion and even provide quantitative estimates of the shear viscosity, they often overlook the fact that strong magnetic fields can substantially modify both the magnitude and the anisotropy of viscous transport coefficients. This limitation partly arises from our still incomplete understanding of how strong magnetic fields influence the viscous properties of relativistic plasmas. Foundational progress in this direction has been made in Refs.~\cite{Huang:2011dc,Critelli:2014kra,Finazzo:2016mhm,Grozdanov:2016tdf,Hattori:2017qih,Hernandez:2017mch,Hattori:2022hyo}, where key aspects of relativistic magnetohydrodynamics were developed recently. In particular, Ref.~\cite{Hattori:2017qih} analyzed how strong magnetic fields modify the viscous coefficients of the QGP and concluded that, within the leading-log and lowest Landau level approximations, only the longitudinal component of the bulk viscosity is nontrivial. As we show in this work, however, contributions from higher Landau levels are essential even in extremely strong fields. Once additional levels are included, the transverse bulk viscosity also becomes nonzero and can play a significant role. Viscous effects may influence heavy-ion observables, such as anisotropic flow and charge-dependent correlations \cite{Tuchin:2011jw}, underscoring the importance of consistently incorporating magnetic fields into theoretical descriptions of the QGP.

In this study, we focus on calculating the shear, bulk, and cross viscosity coefficients of strongly magnetized relativistic plasmas from first principles, within the framework of quantum field theory. By applying Kubo's linear response formalism to a magnetized medium, we express these viscosities in terms of the corresponding correlation functions of the energy-momentum tensor. Our goal is to provide a consistent microscopic description of viscous behavior in magnetized relativistic plasmas by going beyond phenomenological approaches \cite{Ghosh:2022xtv,Nam:2013fpa,Dey:2019axu,Ghosh:2018cxb,Mohanty:2018eja,Kurian:2018dbn,Ghosh:2020wqx}, which often rely on simplifying assumptions or omit essential aspects of the underlying microscopic dynamics.

The paper is organized as follows. In Sec.~\ref{sec:Kubo-Formalism}, we begin with the definitions of the various types of viscosities in terms of correlation functions of the corresponding components of the energy-momentum tensor. Using the spectral representation of the fermion propagator in the Landau-level basis, we then express the shear, bulk, and cross viscosities in terms of the damping rates of charge carriers, $\Gamma_{n}(k_{z})$, in Sec.~\ref{sec:viscosities}. The physical interpretation and formal expressions for these damping rates, originally derived in Ref.~\cite{Ghosh:2024hbf}, are briefly reviewed in Sec.~\ref{sec:damping-rate}. The numerical results for the viscosities in a strongly magnetized electron-positron (QED) plasma are presented in Sec.~\ref{sec:viscosityQED}. In Sec.~\ref{sec:viscosityQCD}, we extend the analysis to a strongly magnetized quark-gluon (QCD) plasma. Finally, Sec.~\ref{sec:summary} summarizes our main findings and conclusions. Several technical derivations and supplementary results are provided in the Appendixes at the end of the paper.

\section{Kubo's formalism for viscosities}
\label{sec:Kubo-Formalism}

Within Kubo's formalism, the various components of the viscosity tensor are expressed in terms of the retarded correlation functions of the energy-momentum tensor. For example, a generic component of the viscosity tensor can be written as follows~\cite{Kapusta:2006pm,Critelli:2014kra}:
\begin{equation}
\eta_{ijkl} = \lim_{p_{0} \to 0} \frac{1}{p_{0}} \, \text{Im}\, G^R_{T^{ij},T^{kl}}(p_{0},\bm{0}),
\end{equation}
where $i,j,k,l$ denote spatial indices. For Dirac fermions, the operator for the energy-momentum current is given by the following expression \cite{Goedecke:1974}:
\begin{equation}
T^{\mu \nu}  = \frac{i}{4}\left[ \bar{\psi} \gamma^{\mu} ({\cal D}^{\nu} \psi ) - ({\cal D}^{\nu} \bar{\psi}) \gamma^{\mu}\psi+\bar{\psi} \gamma^{\nu} ({\cal D}^{\mu}\psi) - ({\cal D}^{\mu}\bar{\psi}) \gamma^{\nu} \psi\right] .
\label{def-T-mu-nu}
\end{equation}
where ${\cal D}^{\mu}\equiv \partial^{\mu} + ie_{f} A^{\mu}$ is the covariant derivative with the vector potential $A_{\rm ext}^{\mu}$ that descries the background magnetic field. Without the loss of generality, we assume that  the background gauge field is $\bm{A}=(0,Bx,0)$. For simplicity, we consider a QED-like plasma containing $N_f$ fermion species with electric charges $e_f$ and mass $m$. The simplest electron-positron plasma corresponds to the single-flavor case with $e_f =-e$ and $m\approx 0.511~\mbox{MeV}$. The generalization to a multi-component system is straightforward.

In an isotropic plasma, the viscosity tensor reduces to the form
\begin{eqnarray}
\eta_{ijkl} = \eta \left(\delta_{ik}\delta_{jl} + \delta_{il}\delta_{jk} - \tfrac{2}{3}\delta_{ij}\delta_{kl}\right)
+ \zeta\, \delta_{ij}\delta_{kl},
\end{eqnarray}
where the scalar coefficients $\eta$ and $\zeta$ represent the shear and bulk viscosities, respectively.

In the presence of a background magnetic field, the transport properties of a relativistic plasma become direction-dependent due to the explicit breaking of spatial isotropy. Consequently, both the shear and bulk viscosities become anisotropic \cite{Huang:2011dc,Critelli:2014kra,Finazzo:2016mhm}. The shear viscosity tensor, in particular, decomposes into several independent components that distinguish between viscous stresses acting parallel or perpendicular to the magnetic field. In the most general case, it contains up to five independent coefficients, corresponding to different directions of momentum flow relative to the field~\cite{Huang:2009ue,Braginskii1965}.

In the present study, we focus on plasmas at zero chemical potential, where charge-conjugation symmetry is preserved. Under this condition, only two shear viscosity coefficients, parallel and perpendicular, remain nonzero, while the other components vanish. The bulk viscosity similarly splits into longitudinal and transverse parts, which differ in the presence of a magnetic field. In addition, a single cross viscosity, $\zeta_{\times}$, survives even at zero chemical potential. This coefficient characterizes the dissipative transport associated with the coupling between longitudinal stress and transverse pressure (and vice versa) \cite{Hattori:2022hyo}. Strictly speaking, $\zeta_{\times}$ is neither a shear nor a bulk viscosity, but rather an independent transport coefficient specific to magnetized plasmas.

It is worth noting that in Ref.~\cite{Huang:2011dc}, $\zeta_{\times}$ was not used as one of the independent transport coefficients. Instead, it appeared as a linear combination of two shear viscosities, $\zeta_{\times}=\frac{1}{2}\eta^{HSR}_{1} + \frac{2}{3}\eta^{HSR}_{0}$. In contrast, Ref.~\cite{Hattori:2022hyo} treated $\zeta_{\times}$ as an independent coefficient, grouping it together with the bulk viscosities. In the present work, we adopt a convention similar to that of Ref.~\cite{Huang:2011dc} for the transverse ($\eta_{\perp}$, $\zeta_{\perp}$) and longitudinal ($\eta_{\parallel}$, $\zeta_{\parallel}$) viscosity components, as these possess well-defined limiting values that coincide with those of an isotropic plasma at $B = 0$. Strictly speaking, the authors of Ref.~\cite{Huang:2011dc} used a different notation for the shear viscosities, namely $\eta^{HSR}_{0}$ and $\eta^{HSR}_{2}$, whose relation to the conventional transverse and longitudinal components reads \cite{Finazzo:2016mhm}: $\eta_{\perp} = \eta^{HSR}_{0}$ and $\eta_{\parallel} = \eta^{HSR}_{0} - \eta^{HSR}_{2}$.

For the cross viscosity $\zeta_{\times}$, however, we follow the definition of Ref.~\cite{Hattori:2022hyo}, as it provides a simpler formulation than the one involving $\eta^{HSR}_{1}$ in Ref.~\cite{Huang:2011dc}. This choice has an important implication: $\zeta_{\times}$ does not vanish in the limit of zero magnetic field. Moreover, it is not an independent transport coefficient in this limit, but instead reduces to a linear combination of the shear and bulk viscosities, $\zeta_{\times}^{(B\to 0)} = \zeta_{0} - \frac{2}{3}\eta_{0}$. In a magnetized plasma, on the other hand, $\zeta_{\times}$ becomes a genuinely new and independent transport characteristics.

In this work, we focus exclusively on the partial contributions of charged fermions to the viscosity coefficients. In principle, one should also account for the contribution from gauge bosons (i.e., medium-modified photons). A parametric estimate for the shear viscosity can be obtained from general considerations \cite{Hattori:2017qih}:
\begin{equation}
\eta_\gamma \sim \frac{T^4}{\Gamma_\gamma} \sim \frac{T^5}{\alpha |eB|} ,
\label{eta-photon}
\end{equation}
where we used the photon damping rate $\Gamma_\gamma\sim \alpha |eB|/T$, arising from the dominant $1\leftrightarrow 2$ processes in the strong-field regime \cite{Wang:2021ebh}. Although the contribution in Eq.~(\ref{eta-photon}) is parametrically suppressed, we will show that it is not always negligible when compared with the charged-fermion contributions. In fact, it is likely to be small relative to the fermionic contribution to the longitudinal shear viscosity, but not to the transverse one.

By repeating the arguments of Ref.~\cite{Hattori:2017qih}, one may conclude that the gauge-boson contribution to the bulk viscosity should be also  parametrically suppressed. Although we do not reexamine this issue in detail here, we note that our analysis of the fermion damping rate differs substantially from that of Ref.~\cite{Hattori:2017qih}, especially in the chiral limit. A careful assessment of the relative importance of charged-fermion and gauge-boson contributions therefore remains an open question. In the present work, we address the former in detail, but leave the latter for future study.

\subsection{Shear viscosity in a magnetized plasma} 

Generally, the shear viscosity quantifies a fluid's resistance to shear deformations, that is, to relative motion between adjacent layers sliding past one another. In a magnetized plasma, such shear stresses can arise either parallel or perpendicular to the direction of the magnetic field. Consequently, it is natural to introduce two independent shear viscosity coefficients, the transverse and longitudinal components, which are defined through the corresponding Kubo relations~\cite{Huang:2011dc,Critelli:2014kra,Finazzo:2016mhm}:
\begin{eqnarray}
 \eta_{\perp} &=& \lim_{p_{0} \to 0} \frac{1}{p_{0}} \, \text{Im} \, G^R_{T^{xy},T^{xy}}(p_{0},\bm{0}) ,  
 \label{eta-perp-Kubo} \\
 \eta_{\parallel} &=& \lim_{p_{0} \to 0} \frac{1}{p_{0}} \, \text{Im} \, G^R_{T^{xz},T^{xz}}(p_{0},\bm{0}) .
  \label{eta-parallel-Kubo} 
\end{eqnarray}
In a QED-like theory with $N_f$ fermion species with electric charges $e_f$, the corresponding one-loop correlators can be expressed in terms of the fermion propagators:
\begin{eqnarray}
G^R_{T^{xy},T^{xy}}(p_{0},\mathbf{p}) &=& \sum_f \int \frac{d^4k}{(2\pi)^4} \, \text{Tr} \left[ \Gamma_{xy}(k+p) \,\bar{S}^f(k+p ) \, \Gamma_{xy}(k) \, \bar{S}^f(k) \right], 
\label{G-Txy-Txy}\\
G^R_{T^{xz},T^{xz}}(p_{0},\mathbf{p}) &=& \sum_f \int \frac{d^4k}{(2\pi)^4} \, \text{Tr} \left[ \Gamma_{xz}(k+p) \,\bar{S}^f(k+p ) \, \Gamma_{xz}(k) \, \bar{S}^f(k) \right],  
\label{G-Txz-Txz} 
\end{eqnarray}
where $\bar{S}^f(k)$ is the Fourier transform of the translation invariant part of the propagator in a background magnetic field  \cite{Miransky:2015ava}. 

It is worth noting that the derivation of these correctors is most convenient to start in coordinate space, where the vertices associated with the energy-momentum tensor (\ref{def-T-mu-nu}) involve covariant derivatives, and the fermion propagators contain the well-known Schwinger phases. Upon simplifying the resulting expressions, the Schwinger phases cancel out. Subsequently, by transforming to momentum space in the translation-invariant parts of the propagators, one obtains the results in Eqs.~(\ref{G-Txy-Txy}) and (\ref{G-Txz-Txz}), where the tree-level vertex functions are given by:
\begin{eqnarray}
\Gamma_{ij}(k) = \frac{1}{2}( k_i \gamma_j + k_j \gamma_i ) .
\end{eqnarray}
By substituting these vertices into Eqs.~(\ref{G-Txy-Txy})  and (\ref{G-Txz-Txz}), we derive explicit expressions for the correlators that define the two shear viscosity coefficients:
\begin{eqnarray}
G^R_{T^{xy},T^{xy}}(p_{0},\mathbf{p}) &=& \frac{1}{2} \sum_f\int \frac{d^4k}{(2\pi)^4} \Big\{ (k_{y}+p_{y}) k_{y}\text{Tr} \left[ \gamma_{1} \,\bar{S}^f(k+p ) \, \gamma_{1} \, \bar{S}^f(k) \right] 
+(k_{x}+p_{x}) k_{y}\text{Tr} \left[ \gamma_{2} \,\bar{S}^f(k+p ) \, \gamma_{1} \, \bar{S}^f(k) \right] \Big\}, 
\label{G-Txy-Txy-long}  \\
G^R_{T^{xz},T^{xz}}(p_{0},\mathbf{p}) &=& \frac{1}{4} \sum_f\int \frac{d^4k}{(2\pi)^4} \Big\{ (k_{x}+p_{x}) k_{x} \text{Tr} \left[ \gamma_{3} \,\bar{S}^f(k+p ) \, \gamma_{3} \, \bar{S}^f(k) \right] + (k_{x}+p_{x}) k_{z} \text{Tr} \left[ \gamma_{3} \,\bar{S}^f(k+p ) \, \gamma_{1} \, \bar{S}^f(k) \right] \nonumber\\
&& +  (k_{z}+p_{z}) k_{x} \text{Tr} \left[ \gamma_{1} \,\bar{S}^f(k+p ) \, \gamma_{3} \, \bar{S}^f(k) \right]
+ (k_{z}+p_{z}) k_{z} \text{Tr} \left[ \gamma_{1} \,\bar{S}^f(k+p ) \, \gamma_{1} \, \bar{S}^f(k) \right] \Big\}.
\label{G-Txz-Txz-long} 
\end{eqnarray}
The correlation functions at finite temperature are evaluated within the Matsubara imaginary-time formalism. In this framework, one performs an analytic continuation to Euclidean space by substituting $k_{0} \to i\omega_{k}$ and $p_{0} \to i\Omega_{l}$, where the fermionic and bosonic Matsubara frequencies are given by $\omega_{k} \equiv (2k + 1)\pi T $ and $\Omega_{l} \equiv 2l\pi T $, respectively. Additionally, the energy integration in Minkowski space is replaced by a discrete sum over Matsubara frequencies, 
\begin{equation}
\int\frac{dk_{0}}{2\pi} f(k_{0},p_{0}) \to  T \sum_{k=-\infty}^{\infty} f(i\omega_{k} , i\Omega_{l}).
\label{integral-to-Matsubara-sum}
\end{equation}
After performing the sum, the results are analytically continued back to Minkowski space through the replacement $i\Omega_{l} \to p_{0} +i\epsilon$.

\subsection{Bulk viscosity in a magnetized plasma} 

The bulk viscosity characterizes a fluid's resistance to uniform compression or expansion. In the presence of a background magnetic field, such deformations can occur either along or perpendicular to the field direction. Accordingly, one can define distinct bulk viscosity coefficients, whose values are determined through the following Kubo relations~\cite{Huang:2011dc}:
\begin{eqnarray}
\zeta_{\perp} &=& \lim_{p_{0} \to 0} \frac{1}{3p_{0}} \, \text{Im} \left[ 2 \, G^R_{\tilde{P}_{\perp} \tilde{P}_{\perp}}(p_{0},\bm{0}) + G^R_{\tilde{P}_{\parallel} \tilde{P}_{\perp}}(p_{0},\bm{0}) \right] ,
 \label{zeta-perp-Kubo}\\
  \zeta_{\parallel} &=& \lim_{p_{0} \to 0} \frac{1}{3p_{0}} \, \text{Im} \left[ 2  \, G^R_{\tilde{P}_{\perp} \tilde{P}_{\parallel}}(p_{0},\bm{0}) +  G^R_{\tilde{P}_{\parallel} \tilde{P}_{\parallel}}(p_{0},\bm{0})  \right].  
  \label{zeta-parallel-Kubo}  
  \end{eqnarray}
The cross viscosity is defined by \cite{Hattori:2022hyo}
\begin{eqnarray}
\zeta_{\times} &=& \lim_{p_{0} \to 0} \frac{1}{p_{0}} \, \text{Im} \left[ G^R_{\tilde{P}_{\perp} \tilde{P}_{\parallel}}(p_{0},\bm{0})  \right].  
  \label{zeta-cross-Kubo}  
\end{eqnarray}
The projected pressure operators $\tilde{P}_{\parallel}$ and $ \tilde{P}_{\perp} $ account for the nontrivial thermodynamic derivatives that arise in the presence of a magnetic field. They are defined as 
\begin{eqnarray}
\tilde{P}_{\parallel} &\equiv& P_{\parallel} - v_{s,\parallel}^2 \, \epsilon =T^{33}-v_{s,\parallel}^2 T^{00} , 
\label{P-parallel}
\\
\tilde{P}_{\perp} &\equiv& P_{\perp} - v_{s,\perp}^2 \, \epsilon=\frac{1}{2}(T^{11}+T^{22})- v_{s,\perp}^2 T^{00},
\label{P-perp}
\end{eqnarray}
where $\epsilon \equiv T^{00}$ is the energy density, and $v_{s,\parallel}$ and $v_{s,\perp}$ are the transverse and longitudinal values of the speed of sound. The latter are defined by partial derivatives of pressure with respect to the energy density, i.e.,
\begin{eqnarray}
v_{s,\parallel}^2  &\equiv& \left( \frac{\partial P_{\parallel}}{\partial \epsilon} \right)_B ,
\label{v-squared-parallel}  \\
v_{s,\perp}^2  &\equiv& \left( \frac{\partial P_{\perp}}{\partial \epsilon} \right)_B = v_{s,\parallel}^2 -B \left( \frac{\partial M}{\partial \epsilon} \right)_B .
\label{v-squared-perp}
\end{eqnarray}
In the last relation, $M$ denoted the magnetization. To leading order in coupling the thermodynamic expressions for the transverse and longitudinal pressure, as well as the magnetization are given in Appendix~\ref{app:Thermodynamics}.

It is worth noting that the definition of the transverse pressure, $P_{\perp} =P_{\parallel} -M B$, adopted here and in most magnetohydrodynamic studies, implicitly assumes that the plasma undergoes compression or expansion under a conserved magnetic flux rather than at a fixed magnetic field strength, see Ref.~\cite{Bali:2014kia} for a clear discussion of this issue. This assumption follows from the Alfv\'{e}n theorem, according to which magnetic field lines are frozen into a perfectly conducting plasma and thus move together with the fluid. However, this is self-consistent when the transverse electrical conductivity is sufficiently large, and the magnetic flux through any fluid element remains approximately constant during its motion. It should be noted, however, that the transverse conductivity could be strongly suppressed by a sufficiently strong background magnetic field \cite{Ghosh:2024owm,Ghosh:2024fkg}. Then, the flux-conserving condition may start to  fail and one may need to switch to the other limiting regime with a fixed magnetic field.

The explicit expressions for the three different correlators that enter the definitions of the bulk viscosity coefficients are:
\begin{eqnarray}
 G_{\tilde{P}_{\perp} \tilde{P}_{\parallel}}(p_{0},\mathbf{p}) &=&  \sum_f\int \frac{d^4k}{(2\pi)^4} \, \text{Tr} \left[ \left( \frac{1}{2} (\bm{k}_{\perp}+\bm{p}_{\perp})\cdot\bm{\gamma}_{\perp}  -  v_{s,\perp}^2 (k_{0}+p_{0}) \gamma_{0} \right) \bar{S}^f(k+p) \left(k_{z}\gamma_{3} -  v_{s,\parallel}^2 k_{0} \gamma_{0}  \right)   \bar{S}^f(k) \right], 
\label{G-perp-parallel-long}  \\
 G_{\tilde{P}_{\parallel} \tilde{P}_{\parallel}}(p_{0},\mathbf{p}) &=&  \sum_f\int \frac{d^4k}{(2\pi)^4} \, \text{Tr} \left[ \left( (k_{z}+p_{z})\gamma_{3} -  v_{s,\parallel}^2 (k_{0}+p_{0}) \gamma_{0}  \right) \bar{S}^f(k+p) \left(k_{z}\gamma_{3} -  v_{s,\parallel}^2 k_{0} \gamma_{0}  \right)   \bar{S}^f(k) \right],  
 \label{G-parallel-parallel-long}  \\
 G_{\tilde{P}_{\perp} \tilde{P}_{\perp}}(p_{0},\mathbf{p}) &=&  \sum_f \int \frac{d^4k}{(2\pi)^4} \, \text{Tr} \left[ \left( \frac{1}{2} (\bm{k}_{\perp}+\bm{p}_{\perp})\cdot\bm{\gamma}_{\perp}  - v_{s,\perp}^2 (k_{0}+p_{0}) \gamma_{0}  \right) \bar{S}^f(k+p) \left( \frac{1}{2} \bm{k}_{\perp}\cdot\bm{\gamma}_{\perp}  -  v_{s,\perp}^2 k_{0}\gamma_{0} \right)  \bar{S}^f(k) \right].  
  \label{G-perp-perp-long}
\end{eqnarray}

\section{Analytical expressions for shear, bulk and cross viscosities}
\label{sec:viscosities}

By employing the formal definitions of the viscosity coefficients introduced in the previous section, we now proceed to derive their analytical expressions. For this purpose, it is convenient to make use of the spectral representation of the retarded and advanced fermion propagators, given by
\begin{equation}
 \bar{S}^{f}(k_{0} \pm i 0 ,\bm{k})  = \lim_{\epsilon\to 0} \int_{-\infty}^{\infty} \frac{d k_{0}^\prime A^{f}_{\bm{k}} (k_{0}^\prime) }{k_{0}-k_{0}^\prime+\mu_{f} \pm i\epsilon},  
 \label{prop-spectral-fun}
\end{equation}
where $A_{\bm{k}} (k_{0})$ denotes the spectral function and $\mu_{f}$ is the chemical potential associated with fermions of flavor $f$. (In what follows, we set $\mu_{f} = 0$.) To simplify the notation, here we omitted the fermion flavor index. 

The spectral function $A_{\bm{k}} (k_{0})$, which encodes information about the quasiparticle damping rates and other in-medium properties of the plasma, is formally defined in terms of the propagators as follows:
\begin{eqnarray}
 A^{f}_{\bm{k}} (k_{0}) =\frac{1}{2\pi i}\left[\bar{S}^{f}(k_{0}-i0,\bm{k}) -\bar{S}^{f}(k_{0}+i0,\bm{k})\right].  
 \label{spectral-density-def}
\end{eqnarray}
In the Landau-level representation, the spectral function takes the following explicit form:
\begin{eqnarray}
 A^{f}_{\bm{k}} (k_{0}) &=& ie^{-k_{\perp}^2\ell^2}\sum_{\lambda=\pm}\sum_{n=0}^{\infty} 
 \frac{(-1)^n}{E_{n}}\Big\{
 \left[E_{n} \gamma^{0} 
 -\lambda  k_{z}\gamma^3+\lambda  m \right] \left[{\cal P}_{+}L_{n}\left(2 k_{\perp}^2\ell^2\right)
 -{\cal P}_{-}L_{n-1}\left(2 k_{\perp}^2\ell^2\right)\right] \nonumber\\
 &+&2\lambda  (\bm{k}_{\perp}\cdot\bm{\gamma}_{\perp}) L_{n-1}^{1} \left(2 k_{\perp}^2 \ell^2\right)
 \Big\}\rho(k_{0}, \lambda E_{n}) ,
 \label{spectral-density}
\end{eqnarray}
where $E_{n}=\sqrt{2n|e_{f}B|+k_{z}^2+m^2}$ is the Landau-level energy, $\ell=1/\sqrt{|e_{f}B|}$ is the magnetic length, ${\cal P}_{\pm}=(1\pm i s_{\perp} \gamma^1\gamma^2)/2$ are the spin-up and spin-down projectors, $s_{\perp} = \mbox{sign}(e_f B)$, and $L_{n}^{\alpha}\left(z\right)$ are the generalized Laguerre polynomials \cite{Gradshteyn:1980}. By convention, here we assume that $L_{-1}^{\alpha}\left(z\right)=0$. The Landau-level spectral width is modeled by function $\rho(k_{0}, \lambda E_{n})$, which is a Lorentzian of width $\Gamma_{n}$, centered at $\lambda E_{n}$, i.e.,
\begin{eqnarray}
 \rho(k_{0}, \lambda E_{n})=\frac{1}{\pi}\frac{\Gamma_{n}}{\left(k_{0} -\lambda E_{n} \right)^2+\Gamma_{n}^2}.
\end{eqnarray}
To simplify the notation, an explicit dependence of $E_{n}$ and $\Gamma_{n}$ on the longitudinal momentum $k_{z}$ was suppressed. 

\subsection{Shear viscosity}
\label{subsec:shear}

By employing Kubo's formulas for the transverse and longitudinal components of the shear viscosity, Eqs.~(\ref{eta-perp-Kubo}) and (\ref{eta-parallel-Kubo}), together with the corresponding correlators given in Eqs.~(\ref{G-Txy-Txy-long}) and (\ref{G-Txz-Txz-long}), and using the spectral representation of the fermion propagators in Eq.~(\ref{prop-spectral-fun}), we obtain the following explicit expressions:
\begin{eqnarray}
 \eta_{\perp} &=& - \frac{1}{64 \pi^2 T} \sum_f \int  d^3 \bm{k}  \int_{-\infty}^{\infty} \frac{dk_{0}}{\cosh^2\frac{k_{0}}{2T}}   \Big\{ k_{y}^2  \text{Tr} \left[ \gamma_{1}  A^f_{\bm{k}} (k_{0}) \gamma_{1}   A^f_{\bm{k}} (k_{0})  \right] 
 +k_{x} k_{y}  \text{Tr} \left[ \gamma_{2}  A^f_{\bm{k}} (k_{0}) \gamma_{1}   A^f_{\bm{k}} (k_{0})  \right]  \Big\},
 \label{shear-perp-AA}
\end{eqnarray}
and
\begin{eqnarray}
 \eta_{\parallel} &=& - \frac{1}{128 \pi^2 T} \sum_f \int  d^3 \bm{k}  \int_{-\infty}^{\infty} \frac{dk_{0}}{\cosh^2\frac{k_{0}}{2T}}  \Big\{k_{x}^2  \text{Tr} \left[ \gamma_{3}  A^f_{\bm{k}} (k_{0}) \gamma_{3}   A^f_{\bm{k}} (k_{0})  \right] 
 +k_{z}^2  \text{Tr} \left[ \gamma_{1}  A^f_{\bm{k}} (k_{0}) \gamma_{1}   A^f_{\bm{k}} (k_{0})  \right] \nonumber\\
 && +2k_{x} k_{z} \text{Tr} \left[ \gamma_{3}  A^f_{\bm{k}} (k_{0}) \gamma_{1}   A^f_{\bm{k}} (k_{0})  \right]  \Big\}.
 \label{shear-parallel-AA}
\end{eqnarray}
In the derivation, we used the following Matsubara sum:
\begin{eqnarray}
 T\sum_{k=-\infty}^{\infty} \frac{1}{\left( i \omega_{k}+i\Omega_{l}-k_{0}\right) \left(i \omega_{k} -k_{0}^\prime \right)}
 =\frac{n_F(k_{0})-n_F(k_{0}^\prime)}{k_{0}-k_{0}^\prime-p_{0}-i\epsilon},
 \label{Matsubara-sum}
\end{eqnarray}
where $n_F(k_{0}) \equiv 1/[\exp(k_{0}/T)+1]$ is the Fermi-Dirac distribution function, $\omega_{k} = (2k+1)\pi T$ is the fermionic  Matsubara frequency, and $\Omega_{l} = 2l \pi T$ is the bosonic Matsubara frequency. The latter is replaced by $p_{0}$ through the standard analytic continuation, $i\Omega_{l} \to p_{0} +i\epsilon$. We also took into account that 
\begin{eqnarray}
\lim_{p_{0}\to 0}\frac{1}{p_{0}}\mbox{Im}\left(\frac{n_F(k_{0})-n_F(k_{0}^\prime)}{k_{0}-k_{0}^\prime-p_{0}-i\epsilon}\right) &=&-\frac{\pi}{4T\cosh^2\frac{k_{0}}{2T}}  \delta (k_{0}-k_{0}^\prime).
\label{limit-relation}
\end{eqnarray}
Using the results for the Dirac traces in Appendix~\ref{app:traces}, we then obtain
\begin{eqnarray}
 \eta_{\perp} &=& -\frac{1}{32 \pi T}  \sum_f \sum_{n,n^\prime=0}^{\infty} (-1)^{n+n^\prime} \sum_{\lambda,\lambda^\prime=\pm} 
 \int_{-\infty}^{\infty} \frac{dk_{z} dk_{0}}{\cosh^2\frac{k_{0}}{2T}} 
 \rho(k_{0}, \lambda E_{n})\rho(k_{0}, \lambda^\prime E_{n^\prime}) 
 \nonumber\\
 &\times&  \int  k_{\perp}^3 dk_{\perp} 
 e^{-2k_{\perp}^2\ell^2}  \left( L_{n}  L_{n^\prime-1}+L_{n^\prime}  L_{n-1}\right) 
 \left(1-\lambda \lambda^{\prime} \frac{k_{z}^2+ m^{2}}{E_{n} E_{n^\prime}}\right)   ,
\end{eqnarray}
and 
\begin{eqnarray}
 \eta_{\parallel} &=& \frac{1}{64 \pi T} \sum_f \sum_{n,n^\prime=0}^{\infty} (-1)^{n+n^\prime}
 \sum_{\lambda,\lambda^\prime=\pm} 
 \int_{-\infty}^{\infty} \frac{d k_{z} dk_{0}}{\cosh^2\frac{k_{0}}{2T}}  
 \rho(k_{0}, \lambda E_{n})\rho(k_{0}, \lambda^\prime E_{n^\prime})
 \nonumber\\
 &\times& \int  k_{\perp} dk_{\perp} e^{-2k_{\perp}^2\ell^2}  \Bigg\{k_{\perp}^2 \left(L_{n}L_{n^\prime}+  L_{n-1}L_{n^\prime-1}\right) 
 \left(1+\lambda \lambda^{\prime} \frac{k_{z}^2- m^{2}}{E_{n}E_{n^\prime}} \right) 
 -8 \lambda \lambda^{\prime} \frac{k_{\perp}^4}{E_{n}E_{n^\prime}} L_{n-1}^{1}  L_{n^{\prime} -1}^{1}  \nonumber\\
 &&
 -2k_{z}^2 \left( L_{n}  L_{n^\prime-1}+L_{n^\prime}  L_{n-1}\right) 
 \left(1-\lambda \lambda^{\prime} \frac{k_{z}^2+ m^{2}}{E_{n} E_{n^\prime}}\right) 
- \frac{4 \lambda\lambda^{\prime} k_{\perp}^2 k_{z}^2}{E_{n} E_{n^\prime}} \left[ \left(L_{n} - L_{n-1} \right)  L_{n^{\prime} -1}^{1}  +\left(L_{n^{\prime}}- L_{n^{\prime}-1} \right)  L_{n-1}^{1}   \right]
 \Bigg\},     
\end{eqnarray}
respectively. It is worth noting that one of the contributions to $\eta_{\perp}$, associated with the trace in Eq.~(\ref{Tbb11}), is proportional to $k_{y}^2 (k_{x}^2 - k_{y}^2)$. This term yields a nonzero contribution upon averaging over the polar angle, resulting in $\tfrac{3}{8}k_{\perp}^4 - \tfrac{1}{8}k_{\perp}^4 = \tfrac{1}{4}k_{\perp}^4$.

To integrate over the transverse momentum $k_{\perp}$, we utilize the table integrals in Appendix~\ref{app:integrals}. The corresponding results read
\begin{eqnarray}
 \eta_{\perp} &=& \frac{1}{128 \pi T \ell^4}  \sum_f \sum_{n,n^\prime=0}^{\infty} \sum_{\lambda,\lambda^\prime=\pm} 
 \int_{-\infty}^{\infty} \frac{dk_{z} dk_{0}}{\cosh^2\frac{k_{0}}{2T}} 
 \rho(k_{0}, \lambda E_{n})\rho(k_{0}, \lambda^\prime E_{n^\prime}) 
 \nonumber\\
 &\times&  \left[ n \delta_{n,n^\prime} + (2n+1)\delta_{n+1,n^\prime} + (n+1) \delta_{n+2,n^\prime} \right]
 \left(1-\lambda \lambda^{\prime} \frac{k_{z}^2+ m^{2}}{E_{n} E_{n^\prime}}\right) ,    
 \label{shear-perp-1}
\end{eqnarray}
and
\begin{eqnarray}
 \eta_{\parallel} &=& \frac{1}{512  \pi T \ell^4} \sum_f \sum_{n,n^\prime=0}^{\infty} 
 \sum_{\lambda,\lambda^\prime=\pm} 
 \int_{-\infty}^{\infty} \frac{d k_{z} dk_{0}}{\cosh^2\frac{k_{0}}{2T}}  
 \rho(k_{0}, \lambda E_{n})\rho(k_{0}, \lambda^\prime E_{n^\prime})
 \nonumber\\
 &\times& \Bigg\{ \left[(4n+\delta_{n,0})\delta_{n,n^\prime} +2(2n+1) \delta_{n+1,n^\prime}    \right]
 \left(1+\lambda \lambda^{\prime} \frac{k_{z}^2- m^{2}}{E_{n}E_{n^\prime}} \right) \nonumber\\
 &&
 - \frac{8 n n^\prime \lambda \lambda^{\prime}}{E_{n}E_{n^\prime} \ell^2}  \left(\delta_{n,n^\prime} + \delta_{n+1,n^\prime}  \right)
 +8k_{z}^2 \ell^2   \delta_{n+1,n^\prime}  
 \left(1-\lambda \lambda^{\prime} \frac{k_{z}^2+ m^{2}}{E_{n} E_{n^\prime}}\right) 
 + \frac{8(n+n^\prime) \lambda\lambda^{\prime} k_{z}^2}{E_{n} E_{n^\prime}} \left(\delta_{n,n^\prime} +\delta_{n+1,n^\prime} \right) 
  \Bigg\},
\label{shear-parallel-1}
\end{eqnarray}
respectively. Finally, performing the sum over $\lambda$ and $\lambda^{\prime}$ using the results in Appendix~\ref{app:sums}, we arrive at the following analytical expression for the transverse shear viscosity:
\begin{eqnarray}
\eta_{\perp} &=& \frac{1}{32  \pi^3 T \ell^4}  \sum_f \sum_{n,n^\prime=0}^{\infty}  
 \int_{-\infty}^{\infty} \frac{dk_{z} dk_{0}}{\cosh^2\frac{k_{0}}{2T}} 
 \frac{\Gamma_{n} \Gamma_{n^\prime} 
 \left[ \left(k_{0}^2+E_{n}^2+\Gamma_{n}^2\right)\left(k_{0}^2+E_{n^\prime}^2+\Gamma_{n^\prime}^2\right)-4k_{0}^2 (k_{z}^2+m^2) \right] }
 {\left[\left(E_{n}^2+\Gamma_{n}^2-k_{0}^2\right)^2+4 k_{0}^2\Gamma_{n}^2\right]\left[\left(E_{n^\prime}^2+\Gamma_{n^\prime}^2-k_{0}^2\right)^2+4 k_{0}^2\Gamma_{n^\prime}^2\right]}
 \nonumber\\
 &\times&  \left[n \delta_{n,n^\prime} + (2n+1)\delta_{n+1,n^\prime} + (n+1) \delta_{n+2,n^\prime}  \right]
\label{eta-perp-final}
\end{eqnarray}
and following longitudinal shear viscosity:
\begin{eqnarray}
 \eta_{\parallel} &=& \frac{1}{128  \pi^3 T \ell^4} \sum_f \sum_{n,n^\prime=0}^{\infty} 
 \int_{-\infty}^{\infty} \frac{d k_{z} dk_{0}}{\cosh^2\frac{k_{0}}{2T}}  
 \frac{\Gamma_{n} \Gamma_{n^\prime} }{\left[\left(E_{n}^2+\Gamma_{n}^2-k_{0}^2\right)^2+4 k_{0}^2\Gamma_{n}^2\right]\left[\left(E_{n^\prime}^2+\Gamma_{n^\prime}^2-k_{0}^2\right)^2+4 k_{0}^2\Gamma_{n^\prime}^2\right]}
 \nonumber\\
 &\times& \Bigg\{ \left[ \delta_{n,0}\delta_{n^\prime,0} +2(n+n^\prime) \left(\delta_{n,n^\prime} +\delta_{n+1,n^\prime}  \right)  \right]
 \left[ \left(k_{0}^2+E_{n}^2+\Gamma_{n}^2\right)\left(k_{0}^2+E_{n^\prime}^2+\Gamma_{n^\prime}^2\right)+4k_{0}^2 (k_{z}^2-m^2) \right] \nonumber\\
 &&
 - \frac{32 n n^\prime k_{0}^2 }{\ell^2}  \left( \delta_{n,n^\prime} +\delta_{n+1,n^\prime}  \right)
 +8k_{z}^2 \ell^2   \delta_{n+1,n^\prime}  
 \left[ \left(k_{0}^2+E_{n}^2+\Gamma_{n}^2\right)\left(k_{0}^2+E_{n^\prime}^2+\Gamma_{n^\prime}^2\right)-4k_{0}^2 (k_{z}^2+m^2)    \right] 
\nonumber\\
 &&+  32(n+n^\prime) k_{0}^2 k_{z}^2 \left(\delta_{n,n^\prime} +\delta_{n+1,n^\prime} \right)
 \Bigg\} . 
 \label{eta-parallel-final}
\end{eqnarray}
These expressions are the main analytical results for the transverse and longitudinal components of the shear viscosity. The explicit form of Eq.~\eqref{eta-parallel-final} shows that the longitudinal shear viscosity receives contributions only from quantum transitions between adjacent Landau levels, $n$ and $n+1$. This behavior resembles that of charge and heat transport \cite{Shovkovy:2025yvn}, which are defined through correlation functions of vector operators. In contrast, the transverse component in Eq.~\eqref{eta-perp-final} receives contributions from both adjacent and second-nearest Landau levels, involving the indices $n$, $n+1$, and $n+2$. This is consistent with expectations, as the corresponding Kubo formula involves a correlation function of second-rank tensor operators.

It is instructive to note that the two components of the shear viscosity given in Eqs.~(\ref{shear-perp-1}) and (\ref{shear-parallel-1}) reduce to the same expression in the limit of a vanishing magnetic field, $\eta_{\perp}^{(B\to0)} =\eta_{\parallel}^{(B\to0)} =\eta_{0}$. This is indeed expected since the plasma becomes isotropic in the absence of the background field. The corresponding limiting case is discussed in Appendix~\ref{app:limit-B0-shear}. In the limit $B\to 0$, the sum over the discrete Landau-level index $n$ is replaced by an integral over the continuous variable $x = 2n|e_f B|$, which plays the role of  the squared transverse momentum.

We use these analytical expressions to obtain the numerical results for hot magnetized QED and QCD plasmas in Secs.~\ref{sec:viscosityQED} and \ref{sec:viscosityQCD}, respectively.

\subsection{Bulk viscosity}
\label{subsec:bulk}

Following the same steps as in the preceding subsection, here we derive analytical expressions for the transverse and longitudinal components of the bulk viscosity. Starting from Kubo's formulas in Eqs.~(\ref{zeta-perp-Kubo}) and (\ref{zeta-parallel-Kubo}), defined in terms of the correlation functions in Eqs.~(\ref{G-perp-parallel-long}) through (\ref{G-perp-perp-long}), and employing the spectral representation of the fermion propagator in Eq.~(\ref{prop-spectral-fun}), we derive the following explicit expressions:
\begin{eqnarray}
\zeta_{\perp} &=& -\frac{1}{96\pi^2 T} \sum_f
 \int \frac{d^3\bm{k} d k_{0}}{\cosh^2\frac{k_{0}}{2T}} \Big[2\text{Tr} \left( \frac{\bm{k}_{\perp}\cdot \bm{\gamma}_{\perp}}{2} A^f_{\bm{k}} (k_{0})\frac{\bm{k}_{\perp}\cdot \bm{\gamma}_{\perp}}{2}   A^f_{\bm{k}} (k_{0}) \right)
 +v_{s,\perp}^2 (v_{s,\parallel}^2+2v_{s,\perp}^2) k_{0}^2 \text{Tr} \left( \gamma_{0} A^f_{\bm{k}} (k_{0}) \gamma_{0}     A^f_{\bm{k}} (k_{0}) \right)  \nonumber\\
&-& (v_{s,\parallel}^2+4v_{s,\perp}^2) k_{0} k_{x} \text{Tr} \left( \gamma_{0} A^f_{\bm{k}} (k_{0}) \gamma_{1} A^f_{\bm{k}} (k_{0}) \right) 
-v_{s,\perp}^2k_{0} k_{z} \text{Tr} \left( \gamma_{0} A^f_{\bm{k}} (k_{0}) \gamma_{3} A^f_{\bm{k}} (k_{0}) \right) 
+k_{x} k_{z} \text{Tr} \left( \gamma_{1} A^f_{\bm{k}} (k_{0}) \gamma_{3} A^f_{\bm{k}} (k_{0}) \right) 
 \Big], \nonumber \\\label{zeta-perp-0}
\end{eqnarray}
and
\begin{eqnarray}
 \zeta_{\parallel} &=& -\frac{1}{96\pi^2 T} \sum_f
 \int \frac{d^3\bm{k} d k_{0}}{\cosh^2\frac{k_{0}}{2T}} \Big[ 
 k_{z}^2 \text{Tr} \left( \gamma_{3} A^f_{\bm{k}} (k_{0}) \gamma_{3}     A^f_{\bm{k}} (k_{0}) \right)
 +v_{s,\parallel}^2 (v_{s,\parallel}^2+2v_{s,\perp}^2) k_{0}^2 \text{Tr} \left( \gamma_{0} A^f_{\bm{k}} (k_{0}) \gamma_{0}     A^f_{\bm{k}} (k_{0}) \right) \nonumber\\
&-&2(v_{s,\parallel}^2+v_{s,\perp}^2) k_{0} k_{z}  \text{Tr} \left( \gamma_{0} A^f_{\bm{k}} (k_{0}) \gamma_{3} A^f_{\bm{k}} (k_{0}) \right)
+2 k_{x} k_{z}\text{Tr} \left( \gamma_{1} A^f_{\bm{k}} (k_{0}) \gamma_{3}  A^f_{\bm{k}} (k_{0}) \right)
-2v_{s,\parallel}^2 k_{0} k_{x} \text{Tr} \left( \gamma_{0} A^f_{\bm{k}} (k_{0}) \gamma_{1}  A^f_{\bm{k}} (k_{0}) \right) 
 \Big], \nonumber\\\label{zeta-para-0}
 \end{eqnarray}
for the transverse and longitudinal components of the bulk viscosity, respectively. In the derivation, we used the same standard Matsubara sum in Eq.~(\ref{Matsubara-sum}) and took into account the relation in Eq.~(\ref{limit-relation}).

Similarly, utilizing Kubo's definition in Eq.~(\ref{zeta-cross-Kubo}) with the correlator in Eq.~(\ref{G-perp-parallel-long}), we derive the expression for the cross viscosity:
\begin{eqnarray}
\zeta_{\times} &=& -\frac{1}{32\pi^2 T} \sum_f
 \int \frac{d^3 \bm{k} d k_{0}}{\cosh^2\frac{k_{0}}{2T}}
\Big[ k_{x} k_{z}\text{Tr} \left( \gamma_{1} A^f_{\bm{k}} (k_{0}) \gamma_{3}  A^f_{\bm{k}} (k_{0}) \right)
+v_{s,\parallel}^2 v_{s,\perp}^2 k_{0}^2 \text{Tr} \left( \gamma_{0} A^f_{\bm{k}} (k_{0}) \gamma_{0} A^f_{\bm{k}} (k_{0}) \right) \nonumber\\
&&
- v_{s,\perp}^2 k_{0} k_{z} \text{Tr} \left( \gamma_{0} A^f_{\bm{k}} (k_{0}) \gamma_{3} A^f_{\bm{k}} (k_{0}) \right)
-v_{s,\parallel}^2 k_{0} k_{x} \text{Tr} \left( \gamma_{0} A^f_{\bm{k}} (k_{0}) \gamma_{1}  A^f_{\bm{k}} (k_{0}) \right) 
\Big] .
\label{zeta-cross-0a}
\end{eqnarray}
Now, utilizing the results for Dirac traces in Appendix~\ref{app:traces}, we derive the following results:
\begin{eqnarray}
 \zeta_{\perp} &=&  \frac{1}{24\pi T} \sum_f \sum_{n,n^\prime=0}^{\infty} (-1)^{n+n^\prime}
 \sum_{\lambda,\lambda^\prime=\pm} 
 \int_{-\infty}^{\infty} \frac{d k_{z} dk_{0}}{\cosh^2\frac{k_{0}}{2T}}  
 \rho(k_{0}, \lambda E_{n})\rho(k_{0}, \lambda^\prime E_{n^\prime})
 \nonumber\\
 &\times& \int  k_{\perp} dk_{\perp} e^{-2k_{\perp}^2\ell^2}  \Bigg\{
 -\frac{k_{\perp}^2}{2} \left( L_{n}  L_{n^\prime-1}+L_{n^\prime}  L_{n-1}\right) 
 \left(1-\lambda \lambda^{\prime} \frac{k_{z}^2+ m^{2}}{E_{n} E_{n^\prime}}\right)
 +4\lambda \lambda^{\prime}  \frac{k_{\perp}^4}{E_{n}E_{n^\prime}} L_{n-1}^{1} L_{n^{\prime}-1}^{1}
  \nonumber\\
 &+&v_{s,\perp}^2   (v_{s,\parallel}^2+2v_{s,\perp}^2)  k_{0}^2 \left(L_{n}L_{n^\prime}+  L_{n-1}L_{n^\prime-1}\right) 
 \left(1+\lambda \lambda^{\prime} \frac{k_{z}^2 +m^{2}}{E_{n}E_{n^\prime}} \right) 
 + 8 \lambda \lambda^{\prime} v_{s,\perp}^2   (v_{s,\parallel}^2+2v_{s,\perp}^2)  \frac{k_{\perp}^2 k_{0}^2}{E_{n}E_{n^\prime}} L_{n-1}^{1}  L_{n^{\prime} -1}^{1}  \nonumber\\
 &+& (v_{s,\parallel}^2+4v_{s,\perp}^2) k_{\perp}^2 \left[\frac{\lambda^{\prime} k_{0}}{E_{n^\prime}} \left(L_{n} - L_{n-1} \right) L_{n^{\prime} -1}^{1} +\frac{\lambda k_{0}}{E_{n}}\left(L_{n^{\prime} } - L_{n^{\prime}-1} \right) L_{n-1}^{1}  \right]\nonumber\\
 &-&v_{s,\perp}^2 k_{z}^2\left(L_{n}L_{n^\prime}+  L_{n-1}L_{n^\prime-1}\right) \left(\frac{\lambda k_{0}}{E_{n}}+\frac{\lambda^\prime k_{0}}{E_{n^\prime}}\right) 
 -\lambda \lambda^\prime \frac{ k_{\perp}^2 k_{z}^2}{E_{n}E_{n^\prime}}\left[ \left(L_{n} - L_{n-1} \right) L_{n^{\prime} -1}^{1} 
 +\left(L_{n^{\prime} } - L_{n^{\prime}-1} \right) L_{n-1}^{1}  \right] 
 \Bigg\}   
\end{eqnarray} 
for the transverse bulks viscosity,
\begin{eqnarray}
 \zeta_{\parallel} &=&  \frac{1}{24\pi T} \sum_f \sum_{n,n^\prime=0}^{\infty} (-1)^{n+n^\prime}
 \sum_{\lambda,\lambda^\prime=\pm} 
 \int_{-\infty}^{\infty} \frac{d k_{z} dk_{0}}{\cosh^2\frac{k_{0}}{2T}}  
 \rho(k_{0}, \lambda E_{n})\rho(k_{0}, \lambda^\prime E_{n^\prime})
 \nonumber\\
 &\times& \int  k_{\perp} dk_{\perp} e^{-2k_{\perp}^2\ell^2}  \Bigg\{k_{z}^2 \left(L_{n}L_{n^\prime}+  L_{n-1}L_{n^\prime-1}\right) 
 \left(1+\lambda \lambda^{\prime} \frac{k_{z}^2- m^{2}}{E_{n}E_{n^\prime}} \right) 
 -8 \lambda \lambda^{\prime} \frac{k_{\perp}^2 k_{z}^2}{E_{n}E_{n^\prime}} L_{n-1}^{1}  L_{n^{\prime} -1}^{1}   \nonumber\\
 &+&v_{s,\parallel}^2 (v_{s,\parallel}^2+2v_{s,\perp}^2)  k_{0}^2 \left(L_{n}L_{n^\prime}+  L_{n-1}L_{n^\prime-1}\right) 
 \left(1+\lambda \lambda^{\prime} \frac{k_{z}^2 +m^{2}}{E_{n}E_{n^\prime}} \right) 
 + 8 \lambda \lambda^{\prime}  v_{s,\parallel}^2  (v_{s,\parallel}^2+2v_{s,\perp}^2)  \frac{k_{\perp}^2 k_{0}^2}{E_{n}E_{n^\prime}} L_{n-1}^{1}  L_{n^{\prime} -1}^{1}  \nonumber\\
 &-&2(v_{s,\parallel}^2+v_{s,\perp}^2) k_{z}^2\left(L_{n}L_{n^\prime}+  L_{n-1}L_{n^\prime-1}\right) \left(\frac{\lambda k_{0}}{E_{n}}+\frac{\lambda^\prime k_{0}}{E_{n^\prime}}\right) 
 -2\lambda \lambda^{\prime}  \frac{k_{\perp}^2 k_{z}^2}{E_{n}E_{n^\prime}}\left[ \left(L_{n} - L_{n-1} \right) L_{n^{\prime} -1}^{1} 
 +\left(L_{n^{\prime} } - L_{n^{\prime}-1} \right) L_{n-1}^{1}  \right] \nonumber\\
 &+&2 v_{s,\parallel}^2 k_{\perp}^2 \left[\frac{\lambda^{\prime} k_{0}}{E_{n^\prime}} \left(L_{n} - L_{n-1} \right) L_{n^{\prime} -1}^{1} +\frac{\lambda k_{0}}{E_{n}}\left(L_{n^{\prime} } - L_{n^{\prime}-1} \right) L_{n-1}^{1}  \right]
 \Bigg\}  
\end{eqnarray} 
for the longitudinal bulks viscosity, and 
\begin{eqnarray}
\zeta_{\times} &=&\frac{1}{8\pi T} \sum_f \sum_{n,n^\prime=0}^{\infty} (-1)^{n+n^\prime}
 \sum_{\lambda,\lambda^\prime=\pm} 
 \int_{-\infty}^{\infty} \frac{d k_{z} dk_{0}}{\cosh^2\frac{k_{0}}{2T}}  
 \rho(k_{0}, \lambda E_{n})\rho(k_{0}, \lambda^\prime E_{n^\prime})
 \nonumber\\
 &\times& \int  k_{\perp} dk_{\perp} e^{-2k_{\perp}^2\ell^2}  \Bigg\{
 -\lambda \lambda^{\prime}  \frac{k_{\perp}^2 k_{z}^2}{E_{n}E_{n^\prime}}\left[ \left(L_{n} - L_{n-1} \right) L_{n^{\prime} -1}^{1} 
 +\left(L_{n^{\prime} } - L_{n^{\prime}-1} \right) L_{n-1}^{1}  \right] \nonumber\\
&&+v_{s,\parallel}^2 v_{s,\perp}^2 k_{0}^2 \left(L_{n}L_{n^\prime}+  L_{n-1}L_{n^\prime-1}\right) 
 \left(1+\lambda \lambda^{\prime} \frac{k_{z}^2 +m^{2}}{E_{n}E_{n^\prime}} \right) 
 + 8 \lambda \lambda^{\prime}  v_{s,\parallel}^2 v_{s,\perp}^2 \frac{k_{\perp}^2 k_{0}^2}{E_{n}E_{n^\prime}} L_{n-1}^{1}  L_{n^{\prime} -1}^{1} \nonumber\\
&&-v_{s,\perp}^2 k_{z}^2 k_{0} \left(L_{n}L_{n^\prime}+  L_{n-1}L_{n^\prime-1}\right) \left(\frac{\lambda}{E_{n}}+\frac{\lambda^\prime}{E_{n^\prime}}\right)  \nonumber\\
&&+v_{s,\parallel}^2  k_{\perp}^2 k_{0} \left[\frac{\lambda^{\prime}}{E_{n^\prime}} \left(L_{n} - L_{n-1} \right) L_{n^{\prime} -1}^{1} +\frac{\lambda}{E_{n}}\left(L_{n^{\prime} } - L_{n^{\prime}-1} \right) L_{n-1}^{1}  \right]
\Bigg\}  ,
\end{eqnarray} 
for the cross viscosity.

Then, integrating over the transverse momentum $k_{\perp}$ using the table integrals in Appendix~\ref{app:integrals}, we obtain
\begin{eqnarray}
 \zeta_{\perp} &=& \frac{1}{96\pi T \ell^2} \sum_f \sum_{n=0}^{\infty}  \beta_{n}
 \sum_{\lambda,\lambda^\prime=\pm} 
 \int_{-\infty}^{\infty} \frac{d k_{z} dk_{0}}{\cosh^2\frac{k_{0}}{2T}}  
 \rho(k_{0}, \lambda E_{n})\rho(k_{0}, \lambda^\prime E_{n}) \Bigg\{
 \frac{n }{4\ell^2} 
 \left(1+2\lambda\lambda^{\prime} + \lambda \lambda^{\prime} \frac{k_{z}^2- 3m^{2}}{E_{n}^2}\right) \nonumber\\
 &+&v_{s,\perp}^2   (v_{s,\parallel}^2+2v_{s,\perp}^2)  k_{0}^2 
 \left(1+\lambda \lambda^{\prime}  \right) 
  -v_{s,\perp}^2   k_{z}^2 k_{0} \frac{\lambda+\lambda^{\prime} }{E_{n}} 
  -(v_{s,\parallel}^2+4v_{s,\perp}^2) \frac{n k_{0}}{2\ell^2} \frac{\lambda^{\prime} +\lambda}{E_{n}}    \Bigg\}\nonumber\\
 &+&
 \frac{1}{96\pi T \ell^4} \sum_f \sum_{n=0}^{\infty}  \sum_{\lambda,\lambda^\prime=\pm} 
 \int_{-\infty}^{\infty} \frac{d k_{z} dk_{0}}{\cosh^2\frac{k_{0}}{2T}}  
 \rho(k_{0}, \lambda E_{n}) \rho(k_{0}, \lambda^\prime E_{n+1})   \bigg[
 \frac{2n+1}{2}  \left(1+\lambda \lambda^{\prime} \frac{k_{z}^2 - m^{2}}{E_{n} E_{n+1}}\right)\nonumber\\
 &-&(v_{s,\parallel}^2+4v_{s,\perp}^2) k_{0} \left(\frac{\lambda^{\prime} (n+1)}{E_{n+1}}+\frac{\lambda n}{E_{n}} \right) 
 +2 \lambda\lambda^{\prime} \frac{n (n+1)}{E_{n} E_{n+1}\ell^2}  
 \bigg] \nonumber\\
 &+&
 \frac{1}{192\pi T \ell^4} \sum_f \sum_{n=0}^{\infty}   \sum_{\lambda,\lambda^\prime=\pm} 
 \int_{-\infty}^{\infty} \frac{d k_{z} dk_{0}}{\cosh^2\frac{k_{0}}{2T}}  
 \rho(k_{0}, \lambda E_{n}) \rho(k_{0}, \lambda^\prime E_{n+2}) (n+1) \left(1-\lambda \lambda^{\prime} \frac{k_{z}^2+ m^{2}}{E_{n} E_{n+2}}\right) ,
\label{zeta-perp-00}
\end{eqnarray}
\begin{eqnarray}
 \zeta_{\parallel} &=&  \frac{1}{96\pi T \ell^2} \sum_f \sum_{n=0}^{\infty} \beta_{n}  
 \sum_{\lambda,\lambda^\prime=\pm} 
 \int_{-\infty}^{\infty} \frac{d k_{z} dk_{0}}{\cosh^2\frac{k_{0}}{2T}}  
 \rho(k_{0}, \lambda E_{n}) \rho(k_{0}, \lambda^\prime E_{n})
 \Bigg[k_{z}^2 
 \left(1+\lambda \lambda^{\prime} \frac{k_{z}^2-m^2}{E_{n}^2 }  \right)  
 +v_{s,\parallel}^2  (v_{s,\parallel}^2+2v_{s,\perp}^2)  k_{0}^2 
 \left(1+\lambda \lambda^{\prime} \right)  \nonumber\\
 &-&2(v_{s,\parallel}^2+v_{s,\perp}^2) k_{z}^2 k_{0} \frac{\lambda +\lambda^{\prime}}{E_{n}} 
 -\frac{n}{\ell^2} v_{s,\parallel}^2k_{0}\frac{\lambda +\lambda^{\prime}}{E_{n}} \Bigg]  \nonumber\\
 &+&  \frac{1}{48\pi T \ell^4} \sum_f \sum_{n=0}^{\infty}  \sum_{\lambda,\lambda^\prime=\pm} 
 \int_{-\infty}^{\infty} \frac{d k_{z} dk_{0}}{\cosh^2\frac{k_{0}}{2T}}  
 \rho(k_{0}, \lambda E_{n}) \rho(k_{0}, \lambda^\prime E_{n+1})  \left[ \frac{(2n+1) \lambda \lambda^{\prime} k_{z}^2}{E_{n} E_{n+1} }
 - v_{s,\parallel}^2 k_{0} \left(\frac{\lambda^{\prime} (n+1)}{E_{n+1}}+\frac{\lambda n}{E_{n}}\right)
 \right] 
\label{zeta-parallel-00}
\end{eqnarray} 
and 
\begin{eqnarray}
\zeta_{\times} &=& \frac{1}{32\pi T\ell^2} \sum_f \sum_{n=0}^{\infty}  \beta_{n} 
 \sum_{\lambda,\lambda^\prime=\pm} \int_{-\infty}^{\infty} \frac{d k_{z} dk_{0}}{\cosh^2\frac{k_{0}}{2T}}  
 \rho(k_{0}, \lambda E_{n})\rho(k_{0}, \lambda^\prime E_{n})\Bigg[
 \frac{n \lambda\lambda^{\prime} k_{z}^2}{E_{n}^2 \ell^2} 
 + v_{s,\parallel}^2  v_{s,\perp}^2  k_{0}^2  \left(1+\lambda \lambda^{\prime} \right)   \nonumber\\
 &&-  v_{s,\perp}^2 k_{z}^2 k_{0} \frac{\lambda +\lambda^{\prime}}{E_{n}}
 -\frac{n}{2\ell^2} v_{s,\parallel}^2  k_{0} \frac{\lambda +\lambda^{\prime}}{E_{n}}
\Bigg]\nonumber\\
 &+&\frac{1}{32 \pi T\ell^4} \sum_f \sum_{n=0}^{\infty} \sum_{\lambda,\lambda^\prime=\pm} \int_{-\infty}^{\infty} \frac{d k_{z} dk_{0}}{\cosh^2\frac{k_{0}}{2T}}  
 \rho(k_{0}, \lambda E_{n})\rho(k_{0}, \lambda^\prime E_{n+1})\Bigg[
 \frac{(2n+1)\lambda\lambda^{\prime} k_{z}^2}{E_{n}E_{n+1}} 
  -v_{s,\parallel}^2  k_{0} \left(\frac{\lambda^{\prime} (n+1)}{E_{n+1}}+\frac{\lambda n}{E_{n}} \right)
\Bigg]
\label{zeta-cross-01}
\end{eqnarray} 
respectively. The spin degeneracy factor is conveniently expressed using the shorthand notation $\beta_{n}=2-\delta_{n,0}$. From the expressions in Eqs.~\eqref{zeta-perp-00} -- \eqref{zeta-cross-01}, we see that the parallel and cross bulk viscosities receive contributions only from transitions between adjacent Landau levels, $n$ and $n+1$, whereas the perpendicular component involves transitions between both adjacent and second-nearest neighboring levels, $n$, $n+1$, and $n+2$.

As in the case of the shear viscosity, one can verify that, in the limit $B\to0$, the transverse and longitudinal components of the bulk viscosity given in Eqs.~(\ref{zeta-perp-00}) and (\ref{zeta-parallel-00}) reduce to the same expression,  $\zeta_{\perp}^{(B\to0)} =\zeta_{\parallel}^{(B\to0)} =\zeta_{0}$. This is shown explicitly in Appendix~\ref{app:limit-B0-bulk}. It is also worth emphasizing that the corresponding zero-field expression in Eq.~(\ref{zeta0}) exhibits the expected behavior in the conformal limit of the plasma, where $v_s^2=1/3$ and $E_{k}=k$, leading to a vanishing bulk viscosity.

In the same limit $B\to0$, the cross viscosity attains a finite negative value, which may appear puzzling at first sight. In fact, as shown in Eq.~(\ref{zeta-cross-0}) in Appendix~\ref{app:limit-B0-cross}, this quantity reduces to a specific linear combination of the shear and bulk viscosities at $B=0$. This behavior is an artifact of the formal definition introduced in Ref.~\cite{Hattori:2022hyo} and adopted in the present work: only in the presence of a background magnetic field does $\zeta_{\times}$ represent a genuinely independent transport coefficient. In light of this observation, one may argue that, rather than using the definition of Ref.~\cite{Hattori:2022hyo}, it could be more physical to construct an alternative linear combination (e.g., $\zeta_{\times} - \zeta_{\perp} +\frac{2}{3}\eta_{\parallel}$) that vanishes in the $B\to0$ limit while remaining finite for $B\neq 0$.

Finally, performing the sum over using the results in Appendix~\ref{app:sums}, we obtain
\begin{eqnarray}
\zeta_{\perp} &=& \frac{1}{24\pi^3 T \ell^2} \sum_f \sum_{n=0}^{\infty}  \beta_{n}
 \int_{-\infty}^{\infty} \frac{d k_{z} dk_{0}}{\cosh^2\frac{k_{0}}{2T}}  
 \frac{ \Gamma_{n}^2}
 {\left[\left(E_{n}^2+\Gamma_{n}^2-k_{0}^2\right)^2+4 k_{0}^2\Gamma_{n}^2\right]^2} 
 \bigg\{ 
\frac{n }{4\ell^2} \left[ \left(k_{0}^2+E_{n}^2+\Gamma_{n}^2\right)^2 +8 k_{0}^2 E_{n}^2
+4  k_{0}^2(k_{z}^2- 3m^{2})\right]  \nonumber\\
 &+&v_{s,\perp}^2   (v_{s,\parallel}^2+2v_{s,\perp}^2)  k_{0}^2 
 \left[ \left(k_{0}^2+E_{n}^2+\Gamma_{n}^2\right)^2 +4k_{0}^2 E_{n}^2\right]
  -4 k_{0}^2 \left[  v_{s,\perp}^2   k_{z}^2  +(v_{s,\parallel}^2+4v_{s,\perp}^2) \frac{n}{2\ell^2}\right]\left(k_{0}^2+E_{n}^2+\Gamma_{n}^2\right)
 \bigg\} \nonumber\\
 &+& \frac{1}{24\pi^3 T \ell^4} \sum_f \sum_{n=0}^{\infty} 
 \int_{-\infty}^{\infty} \frac{d k_{z} dk_{0}}{\cosh^2\frac{k_{0}}{2T}}  
 \frac{ \Gamma_{n}\Gamma_{n+1}}
 {\left[\left(E_{n}^2+\Gamma_{n}^2-k_{0}^2\right)^2+4 k_{0}^2 \Gamma_{n}^2\right]\left[\left(E_{n+1}^2+\Gamma_{n+1}^2-k_{0}^2\right)^2+4 k_{0}^2 \Gamma_{n+1}^2\right] } 
 \nonumber\\
&\times& \bigg\{
\frac{2n+1}{2}  \left[ \left(k_{0}^2+E_{n}^2+\Gamma_{n}^2\right)\left(k_{0}^2+E_{n+1}^2+\Gamma_{n+1}^2\right)+4k_{0}^2 (k_{z}^2-m^2)\right]\nonumber\\
 &-&2 (v_{s,\parallel}^2+4v_{s,\perp}^2) k_{0}^2 \left[n\left(E_{n+1}^2+\Gamma_{n+1}^2+k_{0}^2\right) +(n+1) \left(E_{n}^2+\Gamma_{n}^2+k_{0}^2\right)\right] 
 + 8 k_{0}^2 \frac{n (n+1)}{\ell^2}  
 \bigg\} \nonumber\\
 &+& \frac{1}{48\pi^3 T \ell^4} \sum_f \sum_{n=0}^{\infty}
 \int_{-\infty}^{\infty} \frac{d k_{z} dk_{0}}{\cosh^2\frac{k_{0}}{2T}}  
 \frac{ (n+1)\Gamma_{n}\Gamma_{n+2} \left[
\left(k_{0}^2+E_{n}^2+\Gamma_{n}^2\right)\left(k_{0}^2+E_{n+2}^2+\Gamma_{n+2}^2\right)-4k_{0}^2 (k_{z}^2+m^2)\right] }
 {\left[\left(E_{n}^2+\Gamma_{n}^2-k_{0}^2\right)^2+4 k_{0}^2 \Gamma_{n}^2\right]\left[\left(E_{n+2}^2+\Gamma_{n+2}^2-k_{0}^2\right)^2+4 k_{0}^2 \Gamma_{n+2}^2\right] } , 
\label{zeta-perp-final}
\end{eqnarray} 
\begin{eqnarray}
 \zeta_{\parallel} &=& \frac{1}{24\pi^3 T \ell^2} \sum_f \sum_{n=0}^{\infty} \beta_{n}  
 \int_{-\infty}^{\infty} \frac{d k_{z} dk_{0}}{\cosh^2\frac{k_{0}}{2T}}  
 \frac{\Gamma_{n}^2}{\left[\left(E_{n}^2+\Gamma_{n}^2-k_{0}^2\right)^2+4 k_{0}^2 \Gamma_{n}^2\right]^2} 
 \bigg\{ k_{z}^2\left[  \left(k_{0}^2 +E_{n}^2+\Gamma_{n}^2\right)^2+4 k_{0}^2\left(k_{z}^2-m^2\right)
 \right] \nonumber\\
 &+& v_{s,\parallel}^2  (v_{s,\parallel}^2+2v_{s,\perp}^2)  k_{0}^2 \left[  \left(k_{0}^2 +E_{n}^2+\Gamma_{n}^2\right)^2+4 k_{0}^2 E_{n}^2\right]
 -8k_{0}^2 \left( (v_{s,\parallel}^2+v_{s,\perp}^2)k_{z}^2+v_{s,\parallel}^2\frac{n}{2\ell^2}\right) \left(k_{0}^2 +E_{n}^2+\Gamma_{n}^2\right) \bigg\} \nonumber\\
 &+& \frac{1}{6 \pi^3 T \ell^4}\sum_f \sum_{n=0}^{\infty}  
 \int_{-\infty}^{\infty} \frac{d k_{z} dk_{0}}{\cosh^2\frac{k_{0}}{2T}} 
 \frac{\Gamma_{n}\Gamma_{n+1}k_{0}^2 \left[ 2(2n+1)  k_{z}^2 - v_{s,\parallel}^2 \left[n\left(E_{n+1}^2+\Gamma_{n+1}^2+k_{0}^2\right) +(n+1) \left(E_{n}^2+\Gamma_{n}^2+k_{0}^2\right)\right]\right]}
 {\left[\left(E_{n}^2+\Gamma_{n}^2-k_{0}^2\right)^2+4 k_{0}^2 \Gamma_{n}^2\right]\left[\left(E_{n+1}^2+\Gamma_{n+1}^2-k_{0}^2\right)^2+4 k_{0}^2 \Gamma_{n+1}^2\right] } ,
\label{zeta-parallel-final}
\end{eqnarray} 
and 
\begin{eqnarray}
\zeta_{\times} &=& \frac{1}{32\pi T\ell^2} \sum_f \sum_{n=0}^{\infty}  \beta_{n} 
 \sum_{\lambda,\lambda^\prime=\pm} \int_{-\infty}^{\infty} \frac{d k_{z} dk_{0}}{\cosh^2\frac{k_{0}}{2T}}  
 \rho(k_{0}, \lambda E_{n})\rho(k_{0}, \lambda^\prime E_{n})\Bigg[
 \frac{n \lambda\lambda^{\prime} k_{z}^2}{E_{n}^2 \ell^2} 
 + v_{s,\parallel}^2  v_{s,\perp}^2  k_{0}^2  \left(1+\lambda \lambda^{\prime} \right)   \nonumber\\
 &&-  v_{s,\perp}^2 k_{z}^2 k_{0} \frac{\lambda +\lambda^{\prime}}{E_{n}}
 -\frac{n}{2\ell^2} v_{s,\parallel}^2  k_{0} \frac{\lambda +\lambda^{\prime}}{E_{n}}
\Bigg]\nonumber\\
 &+&\frac{1}{32 \pi T\ell^4} \sum_f \sum_{n=0}^{\infty} \sum_{\lambda,\lambda^\prime=\pm} \int_{-\infty}^{\infty} \frac{d k_{z} dk_{0}}{\cosh^2\frac{k_{0}}{2T}}  
 \rho(k_{0}, \lambda E_{n})\rho(k_{0}, \lambda^\prime E_{n+1})\Bigg[
 \frac{(2n+1)\lambda\lambda^{\prime} k_{z}^2}{E_{n}E_{n+1}} 
  -v_{s,\parallel}^2  k_{0} \left(\frac{\lambda^{\prime} (n+1)}{E_{n+1}}+\frac{\lambda n}{E_{n}} \right)
\Bigg] .
\label{zeta-cross-final}
\end{eqnarray} 
These expressions are the main analytical results for the transverse and longitudinal components of the bulk viscosity, as well as the cross viscosity. We employ them to calculate the numerical results for hot magnetized QED and QCD plasmas in Secs.~\ref{sec:viscosityQED} and \ref{sec:viscosityQCD}, respectively.

\section{Landau-level damping rates}
\label{sec:damping-rate}

To evaluate the shear viscosity coefficients given in Eqs.~(\ref{eta-perp-final}) and (\ref{eta-parallel-final}), the bulk viscosity coefficients in Eqs.~(\ref{zeta-perp-final}) and (\ref{zeta-parallel-final}), and the cross viscosity in Eq.~(\ref{zeta-cross-final}), it is essential to determine the damping rates $\Gamma_{n}(k_{z})$ of the charge carriers. These Landau-level dependent rates determine the widths of the Lorentzian peaks in the fermion spectral function, Eq.~(\ref{spectral-density}), and thus play a crucial role in shaping the transport properties of the system. The explicit expressions for $\Gamma_{n}(k_{z})$ in the Landau-level representation were derived in Ref.~\cite{Ghosh:2024hbf}. For completeness of presentation, we briefly summarize those results for the damping rates in this section.
 
\begin{figure}[t]
 \centering
 \subfigure[]{\includegraphics[width=0.23\textwidth]{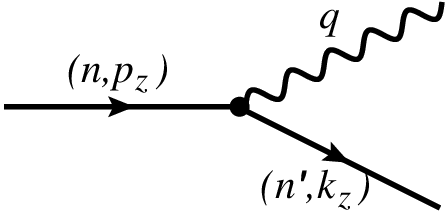}}
 \hspace{0.1\textwidth}
 \subfigure[]{\includegraphics[width=0.23\textwidth]{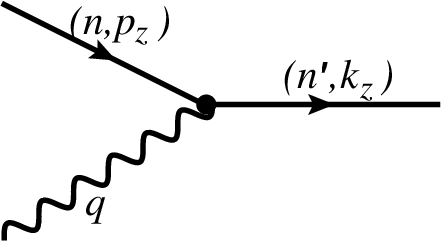}}
 \hspace{0.1\textwidth}
 \subfigure[]{\includegraphics[width=0.23\textwidth]{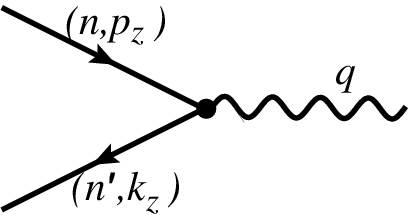}}
 \caption{Leading order processes contributing to the fermion damping rates:
 (a) $\psi_{n}\to \psi_{n^\prime}+\gamma$ with $n >n^{\prime}$, (b) $\psi_{n}+\gamma\to\psi_{n^{\prime}}$ with $n<n^{\prime}$, (c) $\psi_{n}+\bar{\psi}_{n^{\prime}}\to\gamma$, where $n$ and $n^{\prime}$ are the Landau-level indices.}
 \label{Gamma-processes}
\end{figure}

In the regime of a sufficiently strong magnetic field, namely when $|eB| \gg \alpha T^2$, the damping rate $\Gamma_{n}(k_{z})$ of a quantum particle in a given Landau-level state is determined by three classes of processes, illustrated by the Feynman diagrams in Fig.~\ref{Gamma-processes}:
(i) downward transitions to Landau levels with lower indices $n^{\prime}$ ($\psi_{n} \to \psi_{n^{\prime}} + \gamma$ with $n > n^{\prime}$),
(ii) upward transitions to higher Landau levels ($\psi_{n} + \gamma \to \psi_{n^{\prime}}$ with $n < n^{\prime}$), and
(iii) annihilation with negative-energy states ($\psi_{n} + \bar{\psi}_{n^{\prime}} \to \gamma$ for arbitrary $n$ and $n^{\prime}$).

The spin-averaged damping rate in the $n$th Landau level is given by the following expression obtained at the leading order in coupling \cite{Ghosh:2024hbf}:
 \begin{equation}
  \Gamma_{n}(k_{z}) = \frac{\alpha}{2 \ell^2 \beta_{n}  E_{n}} 
  \sum_{n^{\prime}=0}^\infty   \sum_{\{s\}}  
  \int   \frac{d\xi  \left[ 1-n_F(E_{f,s^\prime})+n_B(E_{\gamma,s^\prime}) \right]  {\cal M}_{n,n^{\prime}} (\xi) }{s_{1} s_{2} \sqrt{(\xi-\xi_{n,n^{\prime}}^{-})(\xi-\xi_{n,n^{\prime}}^{+} )} } , 
  \label{Gamma_{n}}
 \end{equation}
where $\xi_{n,n^{\prime}}^{\pm}=\frac{1}{2}\left[\sqrt{2n^{\prime}+(m\ell)^2}\pm \sqrt{2n+(m\ell)^2}\right]^2$ are the dimensionless transverse-momentum threshold parameters. The notation $\sum_{\{s\}}  $ implies summation over three signs, $s^\prime=\pm 1$, $s_{1}=\pm 1$, and $s_{2}=\pm 1$. The function ${\cal M}_{n,n^{\prime}} (\xi)$ appearing in the integrand is determined by the squared amplitude of the leading-order processes shown in Fig.~\ref{Gamma-processes}. Its explicit expression is given by
\begin{eqnarray}
{\cal M}_{n,n^{\prime}}(\xi)  &=&-  \left(n+n^{\prime}+ (m\ell)^2\right)\left[\mathcal{I}_{0}^{n,n^{\prime}}(\xi)+\mathcal{I}_{0}^{n-1,n^{\prime}-1}(\xi) \right] 
+(n+n^{\prime}) \left[\mathcal{I}_{0}^{n,n^{\prime}-1}(\xi)+\mathcal{I}_{0}^{n-1,n^{\prime}}(\xi) \right] ,
  \label{Mnnp}
 \end{eqnarray} 
where the form factor function $\mathcal{I}_{0}^{n,n^{\prime}}(\xi)$ is defined as follows \cite{Wang:2021ebh}: 
 \begin{equation}
  \mathcal{I}_{0}^{n,n^{\prime}}(\xi) =
  \frac{(n^\prime)!}{n!} e^{-\xi}  \xi^{n-n^\prime} \left(L_{n^\prime}^{n-n^\prime}\left(\xi\right)\right)^2   \label{I0}  .
 \end{equation} 
Note that the integration range for $\xi$ in Eq.~(\ref{Gamma_{n}}) depends on the signs of $s_{1}$ and $s_{2}$: (i) for $s_{1}>0$ and any sign of $s_{2}$, the range is $0<\xi<\xi_{n,n^{\prime}}^{-}$ and (ii) for $s_{1}<0$ and $s_{2}>0$, the range is $\xi_{n,n^{\prime}}^{+}<\xi< \infty$. The combination $s_{1} < 0$ and $s_{2} < 0$ does not contribute to the damping rates of particles with positive energy and can be ignored.

The energies of the other fermion ($E_{f,s^\prime}$) and the photon ($E_{\gamma,s^\prime}$) in Eq.~(\ref{Gamma_{n}}) are 
 \begin{eqnarray}
 E_{f,s^\prime}  &=& 
 \frac{E_{n}}{2}\left(1+\frac{2n^{\prime}+(m\ell)^2-2\xi}{2n +(m\ell)^2}  \right)
 + s^\prime k_{z} \frac{\sqrt{(\xi-\xi_{n,n^{\prime}}^{-})(\xi-\xi_{n,n^{\prime}}^{+} )}}{ 2n+(m\ell)^2} ,
 \label{Ef-mass} \\
 E_{\gamma,s^\prime} &=& 
 \frac{E_{n}}{2}\left(1-\frac{2n^{\prime}+(m\ell)^2-2\xi}{2n +(m\ell)^2}  \right)
 - s^\prime k_{z} \frac{\sqrt{(\xi-\xi_{n,n^{\prime}}^{-})(\xi-\xi_{n,n^{\prime}}^{+} )}}{ 2n+(m\ell)^2} ,
 \label{Egamma-mass}
 \end{eqnarray}
respectively. These expressions result from solving the energy conservation conditions for the relevant one-to-two and two-to-one processes.

In the chiral limit ($m = 0$), however, the corresponding energies for the lowest Landau level ($n = 0$) are
 \begin{eqnarray}
 E_{f}  &=& - \frac{(\xi -n^{\prime})^2 +2n^{\prime} k_{z}^2\ell^2 }{2\ell^2 |k_{z}|(\xi-n^{\prime})} , 
 \label{Ef-chiral} \\
 E_{\gamma}  &=&  \frac{(\xi -n^{\prime})^2 +2\xi k_{z}^2\ell^2}{2 \ell^2 |k_{z}| (\xi -n^{\prime})} ,
 \label{Egamma-chiral} 
 \end{eqnarray}
respectively. Unlike Eqs.~(\ref{Ef-mass}) and (\ref{Egamma-mass}), which allow for two solutions labeled by $s^{\prime} = \pm 1$, only a single solution exists for $n = 0$ in the chiral limit.

\section{Numerical result for viscosities in QED plasma}
\label{sec:viscosityQED}

Using the analytical expressions for the shear, bulk, and cross viscosities derived in Sec.~\ref{sec:viscosities}, together with the damping rates obtained in Sec.~\ref{sec:damping-rate}, we now investigate numerically the corresponding transport properties of a magnetized QED plasma composed of electrons and positrons (i.e., a single-flavor plasma with $e_f=-e$), with the electron chemical potential set to zero. Owing to the smallness of the fine-structure constant, $\alpha=e^{2}/(4\pi)\simeq1/137$, the leading-order expression for the damping rate provides a reliable approximation when the magnetic field is sufficiently strong. In particular, requiring that the one-to-two and two-to-one processes shown in Fig.~\ref{Gamma-processes} dominate over the two-to-two scattering channels implies the range of validity $|eB|\gtrsim \alpha T^{2}$.

Our analysis will focus primarily on the ultra-relativistic regime, $m\ll T$ and $m\ll\sqrt{|eB|}$, which is more amenable to a systematic exploration over a broad parameter space. Specifically, we will consider temperatures and magnetic fields in the ranges $15 m\lesssim T\lesssim 85m$ and $(15m)^{2}\lesssim |eB|\lesssim(225m)^{2}$, respectively. We will also demonstrate that, in this regime, the viscosities, when expressed in units of $T^{3}$, obey approximate scaling relations: they are determined by dimensionless functions that depend solely on the ratio $|eB|/T^{2}$. Although these scaling behaviors become exact only in the chiral limit, they remain good approximations throughout the ultra-relativistic domain even for finite fermion mass.

\subsection{Shear viscosity}
\label{subsec:Shear-numerical}

We begin with the numerical analysis of the shear viscosity. The analytical expressions for its transverse and longitudinal components are given in Eqs.~(\ref{eta-perp-final}) and (\ref{eta-parallel-final}). Their evaluation requires detailed knowledge of the damping rates $\Gamma_{n}(k_{z})$ in the Landau-level representation \cite{Ghosh:2024hbf}, as discussed in Sec.~\ref{sec:damping-rate}. Generating these data with sufficient resolution constitutes one of the most computationally costly parts of the calculation, especially in view of the broad ranges of temperature and magnetic field explored here. For each parameter set, we compute $\Gamma_{n}(k_{z})$ for longitudinal momenta up to $k_{z,{\rm max}}=600m$ and for all Landau levels up to $n=50$, while ensuring that contributions from all processes involving levels with indices up to $n^{\prime}=100$ are fully accounted for.

\begin{figure}[t]
\centering
\includegraphics[width=0.47\textwidth]{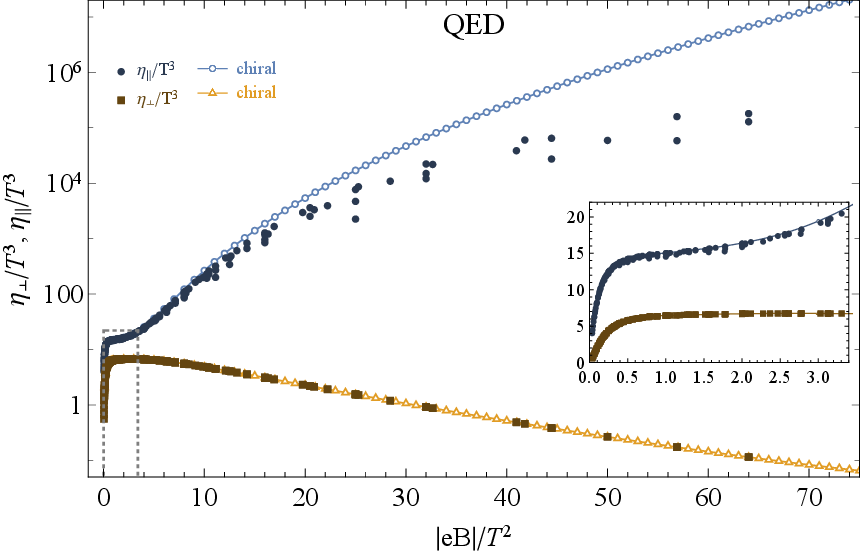}
\hspace{0.03\textwidth}
\includegraphics[width=0.47\textwidth]{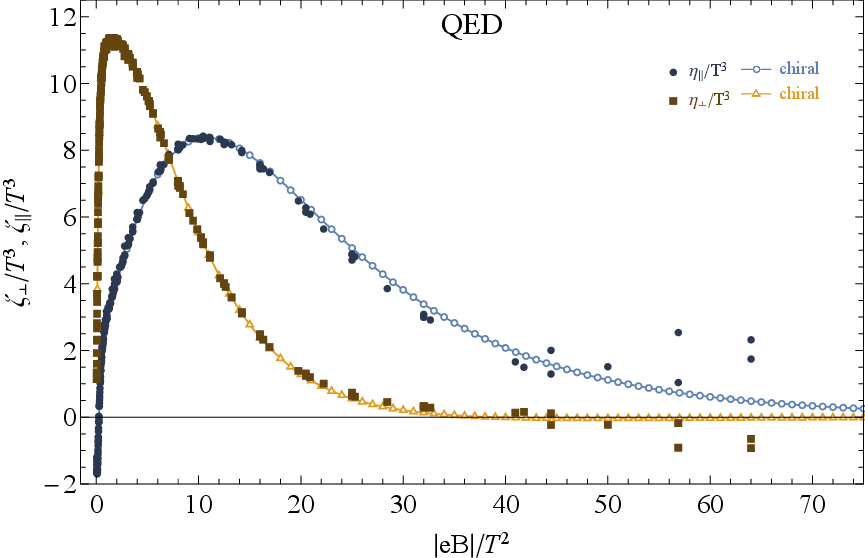} 
\caption{The transverse and longitudinal components of the shear (left) and bulk (right) viscosities as functions of the dimensionless ratio $|eB|/T^2$ in a magnetized QED plasma. Empty markers and interpolating lines represent the results in the chiral limit. Filled markers represents the results for the massive case.}
\label{fig:viscosity-QED}
\end{figure}

The numerical results for the transverse and longitudinal shear viscosities are shown in the left panel of Fig.~\ref{fig:viscosity-QED}. Motivated by the expected scaling behavior, we present the combined data for various temperatures and magnetic-field strengths as functions of the dimensionless ratio $|eB|/T^{2}$. Results for fermions with finite mass are indicated by filled markers, while those in the chiral (massless) limit are shown with open markers. The transverse shear viscosity is represented by the orange and brown points, whereas the light and dark blue points correspond to the longitudinal component.

Observing that most data points collapse onto nearly universal curves, we confirm the anticipated scaling behavior in the ultra-relativistic regime. This behavior is especially pronounced for the transverse shear viscosity, for which the massive and massless results are almost indistinguishable. Interestingly, the transverse component is strongly suppressed by the magnetic field. In contrast, the longitudinal shear viscosity is strongly enhanced by the field. It also exhibits more noticeable departures from scaling, as evidenced by finite-mass effects, which are substantially more pronounced.

In the regime of very strong magnetic fields, $|eB|/T^{2} \gg 1$, it is natural to ask whether the viscosities become dominated by contributions from quantum states in the lowest Landau level. For the longitudinal component, $\eta_{\parallel}$, this is indeed the case. In particular, the term proportional to $\delta_{n,0}\delta_{n^{\prime},0}$ in the integrand of Eq.~(\ref{eta-parallel-final}) gives the leading contribution at large $|eB|$. 
While there are several other terms diagonal in the Landau-level indices, namely those proportional to $n n^{\prime}\delta_{n,n^{\prime}}$ and $(n+n^{\prime})\delta_{n,n^{\prime}}$, they vanish when both $n$ and $n^{\prime}$ are zero. All other nontrivial contributions require either $n\geq 1$ or $n^{\prime}\geq 1$ and, therefore are suppressed. 

The situation is different for the transverse shear viscosity, given by Eq.~(\ref{eta-perp-final}). The only Landau-level-diagonal term in its integrand is proportional to $n\delta_{n,n^{\prime}}$, which vanishes when $n=0$. This feature is ultimately responsible for the strong suppression of the corresponding transport coefficient at large $|eB|$.

We note that the overall behavior of the two shear viscosity components resembles the transverse and longitudinal conductivities studied in Refs.~\cite{Ghosh:2024owm,Ghosh:2024fkg}. Despite this similarity, the underlying mechanisms differ in details. Unlike the charge conductivities, whose simple analytical forms either include or exclude diagonal terms in the Landau-level indices, both viscosities contain such terms. An additional complication arises because some of these contributions are proportional to the Landau-level index $n$, which contribute only when the quantum states are in higher Landau levels ($n>0$). Nevertheless, the rapidly increasing longitudinal component of the shear viscosity is consistent with a conventional transport mechanism, where $\eta_{\parallel} \propto 1/\Gamma_{0}$ with $\Gamma_{0}$ being the damping rate in the lowest Landau level. In contrast, the decreasing transverse shear viscosity signals a qualitatively different mechanism, which requires transitions between the lowest and higher Landau levels, and consequently the corresponding transport coefficient behaves as $\eta_{\perp} \propto \Gamma_{0}$ in the same limit.

Our numerical results show that the transverse shear viscosity becomes sufficiently small to drop below the KSS bound \cite{Kovtun:2004de} at sufficiently strong magnetic fields, specifically when $|eB|/T^2 \gtrsim 40$. This is easy to demonstrate by combining the numerical data for the shear viscosity with the entropy density obtained from Eq.~(\ref{app:entropy}) in Appendix~\ref{app:Thermodynamics}. At first glance, such a result may seem unsurprising, as violations of the bound have been reported previously in the context of strongly coupled anisotropic theories, e.g., see Refs.~\cite{Rebhan:2011vd,Critelli:2014kra,Jain:2015txa,Finazzo:2016mhm,Liu:2016njg}. However, the distinctive feature of the present study is that a similar violation arises within {\em weakly-coupled} QED, albeit only in the strong-field limit.

Before concluding this subsection, it is worth recalling that the numerical calculation does not include the contribution of gauge bosons to the shear viscosity. An order-of-magnitude estimate for such a contribution is given in Eq.~(\ref{eta-photon}). Clearly, it is parametrically suppressed relative to the charged fermion contribution in $\eta_{\parallel}$, but this need not be the case for the transverse component $\eta_{\perp}$. Therefore, a complete determination of $\eta_{\perp}$ requires a more detailed analysis that incorporates both fermionic and bosonic contributions. Nevertheless, the main conclusion remains unchanged: the total transverse shear viscosity is still much smaller than its longitudinal one.

\subsection{Bulk viscosity}
\label{subsec:Bulk-numerical}

Let us now turn to the bulk viscosity of a strongly magnetized QED plasma. The corresponding numerical results for the transverse and longitudinal components are shown in the right panel of Fig.~\ref{fig:viscosity-QED}. As in the case of the shear viscosity, we compile the data for all temperatures and magnetic-field strengths and display them as functions of the dimensionless ratio $|eB|/T^{2}$. 

As seen from Fig.~\ref{fig:viscosity-QED}, the bulk viscosity has several notable features. Both the transverse and longitudinal components exhibit nonmonotonic behavior: each develops a maximum at intermediate field strengths and decreases toward small values in the limits of weak and strong magnetic fields. Surprisingly, the transverse component appears to become negative at large $|eB|/T^{2}$, while the longitudinal component can turn negative at small $|eB|/T^{2}$. Taken at face value, such behavior would signal potential magnetohydrodynamic instabilities.

We do not interpret these negative values as physical. At small $|eB|/T^{2}$, the apparent issues in $\zeta_{\parallel}$ likely stem from the slow convergence of the Landau-level sum as $B\to 0$, which can be compounded by limited resolution in $k_{z}$ used for interpolating the damping rate $\Gamma_{n}(k_{z})$. Likewise, the negative values of $\zeta_{\perp}$ observed for certain model parameters at large $B$ are attributable to numerical limitations, including limited resolution of $\Gamma_{n}(k_{z})$ and extreme sensitivity to input thermodynamic quantities, particularly the speed of sound, rather than to any genuine instability of the plasma.

\begin{figure}[t]
\centering
\includegraphics[width=0.47\textwidth]{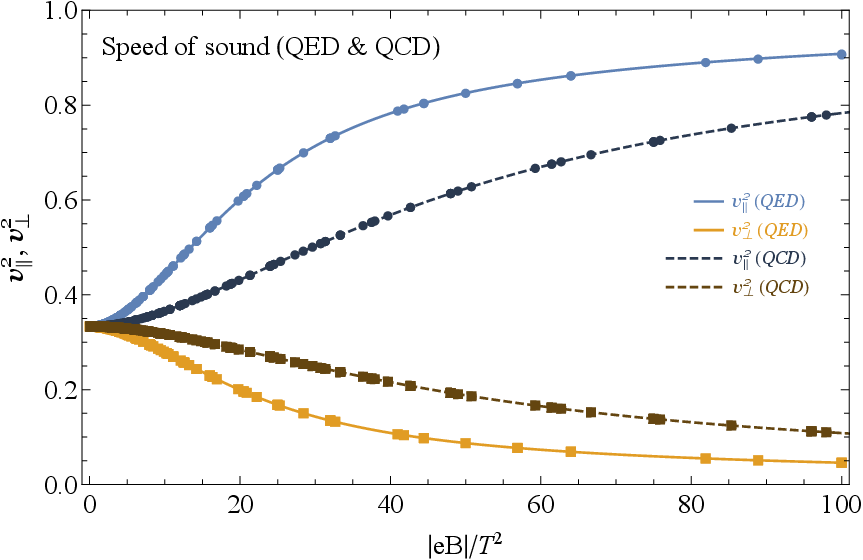}
\hspace{0.03\textwidth}
\includegraphics[width=0.47\textwidth]{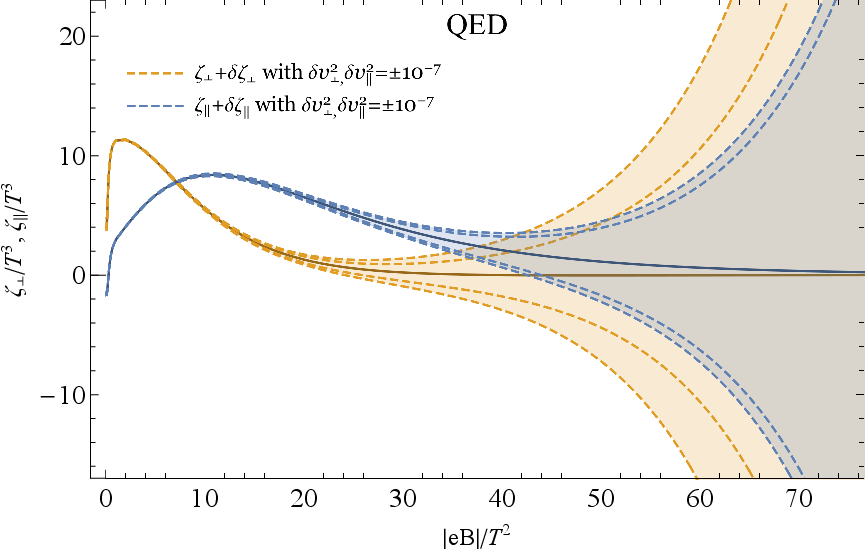}
\caption{Left panel: Transverse and longitudinal components of the speed of sound squared in strongly magnetized QED and QCD plasmas.
Right panel: Transverse and longitudinal bulk viscosities in QED (solid lines), together with their variation range corresponding to $\pm 10^{-7}$ changes in $\delta v_{\perp}^2$ and $\delta v_{\parallel}^2$.}
\label{fig:speed-vars-QED}
\end{figure}

To demonstrate a strong sensitivity of the bulk viscosities to the speed of sound, we examine how the results for $\zeta_{\perp}$ and $\zeta_{\parallel}$ vary when $v_{\perp}^2$ and $v_{\parallel}^2$ are perturbed. From the analytical definitions in Eqs.~(\ref{zeta-perp-final}) and (\ref{zeta-parallel-final}), one finds
\begin{eqnarray}
\delta \zeta_{\perp} &=&  \left[(v_{s,\parallel}^2+4v_{s,\perp}^2) \chi^{(1)} -\chi^{(2)}-4 \chi^{(3)}   \right] \delta v_{s,\perp}^2
+  \left[v_{s,\perp}^2\chi^{(1)} -\chi^{(3)}   \right] \delta v_{s,\parallel}^2
+ \chi^{(1)} \delta v_{s,\parallel}^2\delta v_{s,\perp}^2
+ 2\chi^{(1)} (\delta v_{s,\perp}^2)^2 ,
\label{delta-bulk-perp} \\
\delta \zeta_{\parallel}&=& 2\left(v_{s,\parallel}^2 \chi^{(1)}  -\chi^{(2)}  \right) \delta v_{s,\perp}^2
+  2\left[(v_{s,\perp}^2+v_{s,\parallel}^2) \chi^{(1)} -\chi^{(2)} -\chi^{(3)}   \right] \delta v_{s,\parallel}^2
+ 2\chi^{(1)} \delta v_{s,\parallel}^2\delta v_{s,\perp}^2
+ \chi^{(1)} (\delta v_{s,\parallel}^2)^2,
\label{delta-bulk-para}
\end{eqnarray} 
where the three coefficient functions are defined by 
\begin{eqnarray}
\chi^{(1)} &=&  \frac{1}{24\pi^3 T \ell^2} \sum_f \sum_{n=0}^{\infty}  \beta_{n}
 \int_{-\infty}^{\infty} \frac{d k_{z} dk_{0}}{\cosh^2\frac{k_{0}}{2T}}  
 \frac{ \Gamma_{n}^2  k_{0}^2 
 \left[ \left(k_{0}^2+E_{n}^2+\Gamma_{n}^2\right)^2 +4k_{0}^2 E_{n}^2\right] }
 {\left[\left(E_{n}^2+\Gamma_{n}^2-k_{0}^2\right)^2+4 k_{0}^2\Gamma_{n}^2\right]^2} ,\\
\chi^{(2)} &=&  \frac{1}{6\pi^3 T \ell^2} \sum_f \sum_{n=0}^{\infty}  \beta_{n}
 \int_{-\infty}^{\infty} \frac{d k_{z} dk_{0}}{\cosh^2\frac{k_{0}}{2T}}  
 \frac{k_{0}^2 k_{z}^2 \Gamma_{n}^2 \left(k_{0}^2+E_{n}^2+\Gamma_{n}^2\right) }
 {\left[\left(E_{n}^2+\Gamma_{n}^2-k_{0}^2\right)^2+4 k_{0}^2\Gamma_{n}^2\right]^2},\\
\chi^{(3)} &=&  \frac{1}{12\pi^3 T \ell^4} \sum_f \sum_{n=0}^{\infty}  \beta_{n}
 \int_{-\infty}^{\infty} \frac{d k_{z} dk_{0}}{\cosh^2\frac{k_{0}}{2T}}  
 \frac{ n k_{0}^2 \Gamma_{n}^2 \left(k_{0}^2+E_{n}^2+\Gamma_{n}^2\right) }
 {\left[\left(E_{n}^2+\Gamma_{n}^2-k_{0}^2\right)^2+4 k_{0}^2\Gamma_{n}^2\right]^2} \nonumber \\
&+&   \frac{1}{12\pi^3 T \ell^4} \sum_f \sum_{n=0}^{\infty} 
 \int_{-\infty}^{\infty} \frac{d k_{z} dk_{0}}{\cosh^2\frac{k_{0}}{2T}}  
 \frac{ k_{0}^2 \Gamma_{n}\Gamma_{n+1} \left[n\left(E_{n+1}^2+\Gamma_{n+1}^2+k_{0}^2\right) +(n+1) \left(E_{n}^2+\Gamma_{n}^2+k_{0}^2\right)\right] }
 {\left[\left(E_{n}^2+\Gamma_{n}^2-k_{0}^2\right)^2+4 k_{0}^2 \Gamma_{n}^2\right]\left[\left(E_{n+1}^2+\Gamma_{n+1}^2-k_{0}^2\right)^2+4 k_{0}^2 \Gamma_{n+1}^2\right] } . 
\end{eqnarray} 
The two components of the speed of sound, $v_{\perp}^2$ and $v_{\parallel}^2$, together with the induced variations of the transverse and longitudinal bulk viscosities are shown in Fig.~\ref{fig:speed-vars-QED} as functions of $|eB|/T^{2}$. The shaded regions in the right panel indicate the variation range of the viscosities obtained by varying $\delta v_{\perp}^2$ and $\delta v_{\parallel}^2$  within $\pm 10^{-7}$.

To compute the transverse and longitudinal speeds of sound, $v_{\perp}^2$ and $v_{\parallel}^2$, we employ their thermodynamic definitions given in Eqs.~(\ref{v-perp}) and (\ref{v-parallel}) in Appendix~\ref{app:Thermodynamics}. These expressions are formulated in terms of the magnetization in Eq.~(\ref{app:magnetization}) and the entropy density in Eq.~(\ref{app:entropy}). When evaluating the viscosities at leading order in the coupling, it is formally sufficient to use the thermodynamic functions of the noninteracting (free) theory. Nevertheless, given the strong sensitivity of the viscosities to the precise values of $v_{\perp}^2$ and $v_{\parallel}^2$, interaction effects may play a significant role, unless subtle cancellations of such effects occur at subleading order. A more careful assessment of these effects is left for future work.

The numerical results in the left panel of Fig.~\ref{fig:speed-vars-QED} show that the speed of sound becomes increasingly anisotropic as the magnetic field strength grows. The longitudinal speed $v_{\parallel}$ gradually approaches unity (in units where $c=1$), while $v_{\perp}$  tends to zero. These features can be interpreted as yet another manifestation of the effective dimensional reduction induced by a strong magnetic field. The dependence of $v_{\perp}^2$ and $v_{\parallel}^2$ on the field strength is qualitatively unsurprising and follows straightforwardly from the behavior of the thermodynamic quantities that have been extensively studied in the literature. Yet, to the best of our knowledge, the corresponding results for quark matter have not been reported previously. Most existing studies have focused on nuclear matter \cite{Tews:2018kmu,Reed:2019ezm,Kanakis-Pegios:2020jnf,Mondal:2023baz,Parui:2025zlb}, with only a few addressing quark matter at nonzero baryon density~\cite{Khaidukov:2020vmz,Ferrer:2022afu}.

As seen from the right panel of Fig.~\ref{fig:speed-vars-QED}, the variations in $\zeta_{\perp}$ and $\zeta_{\parallel}$ grow rapidly with increasing $|eB|/T^{2}$, even as the bulk viscosities themselves decrease. This behavior signals that the results become increasingly unreliable at large $|eB|/T^{2}$. We also note that interaction effects in the thermodynamic relations were not included when computing for $v_{\perp}^2$ and $v_{\parallel}^2$. Given the extreme sensitivity of the bulk viscosities, it is plausible that such interaction effects may play a non-negligible role.

We interpret the overall dependence of the bulk viscosities on the magnetic field as follows. In the limit $B \to 0$, the effect of the weak magnetic field becomes negligible, and both $\zeta_{\perp}$ and $\zeta_{\parallel}$ approach small values. This behavior may indicate an approximate restoration of conformal symmetry in the ultrarelativistic plasma. As the magnetic field increases, the additional scale introduced by $|eB|$ explicitly breaks scale invariance, which in turn leads to an enhancement of the bulk viscosity. In the opposite limit of very strong magnetic fields, however, when the low-energy dynamics of the plasma is predominantly governed by the states in the lowest Landau level, a different form of scale invariance may emerge. This symmetry would be associated with the effectively $1+1$-dimensional longitudinal subspace in the $t$-$z$ directions, reflecting the dimensional reduction associated with the lowest-Landau-level dynamics.

\begin{figure}[t]
\centering
\includegraphics[width=0.47\textwidth]{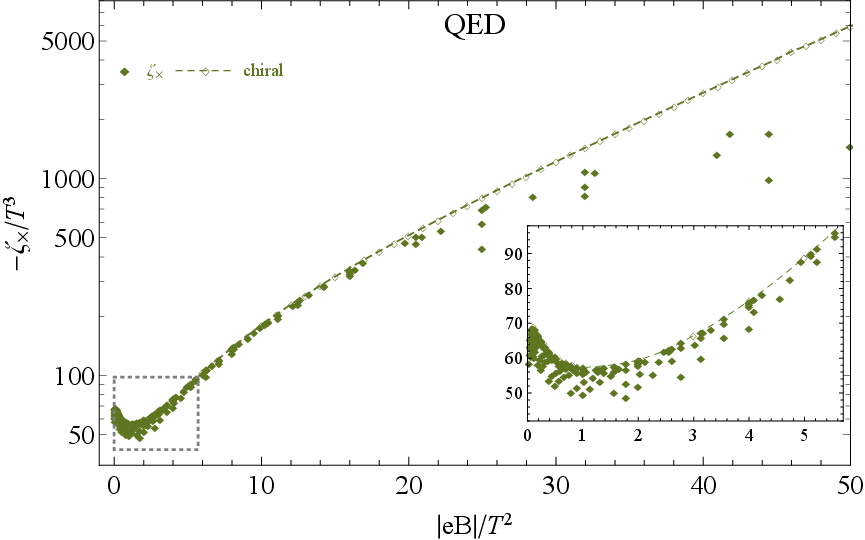}
\hspace{0.03\textwidth}
\includegraphics[width=0.47\textwidth]{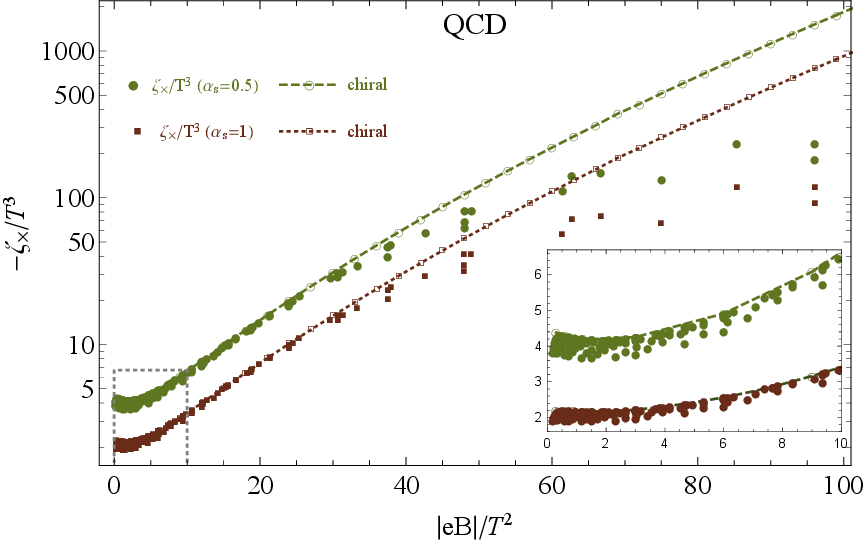} 
\caption{The negative cross viscosity as a function of the dimensionless ratio $|eB|/T^2$ in a magnetized QED (left) and QCD (right) plasmas. Empty markers with interpolating lines correspond to results in the chiral limit, while filled markers represent the massive case.}
\label{fig:cross-viscosity-log}
\end{figure}

For completeness, we also perform a numerical analysis of the cross viscosity defined in Eq.~(\ref{zeta-cross-final}). The corresponding results for finite- and zero-mass QED plasma are shown in the left panel of Fig.~\ref{fig:cross-viscosity-log}. Following the convention of Ref.~\cite{Hattori:2022hyo}, this quantity is defined with an overall negative sign. Accordingly, we plot $-\zeta_{\times}$, which exhibits a nonmonotonic dependence on $|eB|/T^2$, featuring a minimum at values of $|eB|/T^2$ of order unity. At larger magnetic fields, $-\zeta_{\times}$ increases rapidly, in sharp contrast to the other two components of the bulk viscosity, which tend to approach zero in the same limit.

It is instructive to recall that the bulk viscosities must satisfy the inequality $\tilde{\zeta}_{\perp}\tilde{\zeta}_{\parallel} -\zeta_{\times}^2\geq 0$,  
where the transport coefficients $\tilde{\zeta}_{\perp}$ and $\tilde{\zeta}_{\parallel} $ are defined according to the convention of Ref.~\cite{Hattori:2022hyo}. These differ from the definitions employed in Ref.~\cite{Huang:2011dc}, which we use in this work. The relation between the two sets of coefficients is
\begin{eqnarray}
\tilde{\zeta}_{\perp} &=& \frac{3}{2}\zeta_{\perp} -\frac{1}{2} \zeta_{\times},\\
\tilde{\zeta}_{\parallel} &=& 3\zeta_{\parallel} -2 \zeta_{\times} .
\end{eqnarray}
Following our custom conventions adopted in this paper, therefore, the viscosities should satisfy the following inequality: 
\begin{equation}
3\zeta_{\perp}\zeta_{\parallel}  - \left(\zeta_{\parallel} +2 \zeta_{\perp} \right) \zeta_{\times} \geq 0.
\end{equation}
Since $\zeta_{\times}$ is negative, this condition is automatically satisfied.

\section{Numerical result for viscosities in QCD plasma}
\label{sec:viscosityQCD}
 
 The validity of the magnetoviscosity results obtained in this work relies heavily on the strong-field and weak-coupling approximations. The weak-coupling assumption, in particular, becomes a substantial limitation when attempting to extend the analysis to QCD. Although QCD also possesses a weakly-coupled regime, it is realized only at extremely high temperatures, of order the electroweak scale or above. Such conditions are far from those achieved in the quark-gluon plasma created in heavy-ion collisions, where understanding the viscous properties is of great phenomenological interest. In this experimentally relevant regime, the coupling constant is of order unity. Consequently, any quantitative extrapolation of our results to a strongly coupled quark-gluon plasma is unlikely to be reliable. Nevertheless, we will explore how the results are modified under these conditions. Our goal is to extract at least qualitative insights and, perhaps, provide a useful benchmark for future studies.
 
Here we consider a hot two-flavor quark-gluon plasma. For simplicity, we assume that the up and down quarks have equal masses, $m = 5~\mbox{MeV}$. It is, however, essential to retain their different physical electric charges, $e_u = 2e/3$ and $e_d = -e/3$, which determine their different couplings to the external magnetic field. One must also account for the fact that each quark flavor comes in $N_c=3$ colors. Aside from these differences, we employ the same sets of temperature and magnetic-field values as in the QED analysis, now expressed in units of the quark mass rather than the electron mass.

The relevant one-to-two and two-to-one processes, illustrated in Fig.~\ref{Gamma-processes}, have direct QCD counterparts obtained by replacing the external photons with gluons. Numerically, this substitution implies that the fine-structure constant in the damping rates must be replaced by $\alpha_s C_F$, where $\alpha_s$ is the strong coupling constant and $C_F = (N_c^2 - 1)/(2N_c)$ is the quadratic Casimir invariant of the fundamental representation. For $N_c = 3$, one has $C_F = 4/3$.

\begin{figure}[t]
\centering
\includegraphics[width=0.47\textwidth]{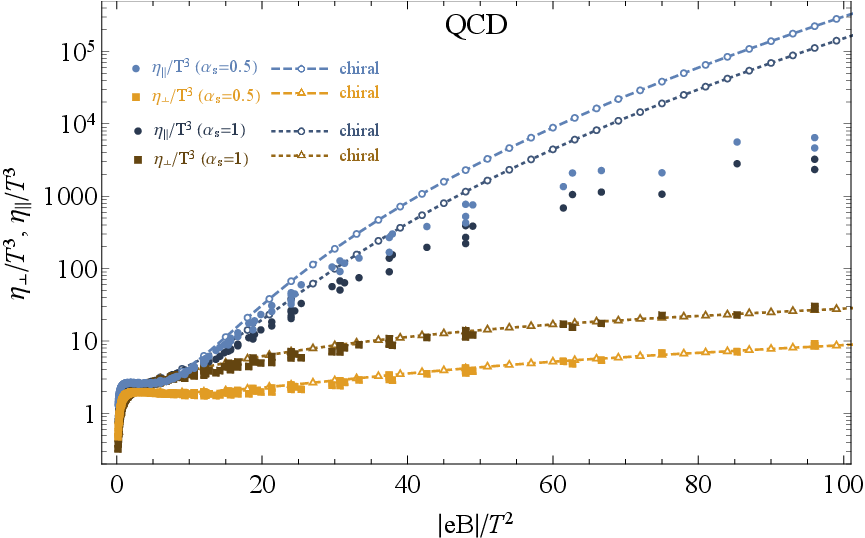}
\hspace{0.03\textwidth}
\includegraphics[width=0.47\textwidth]{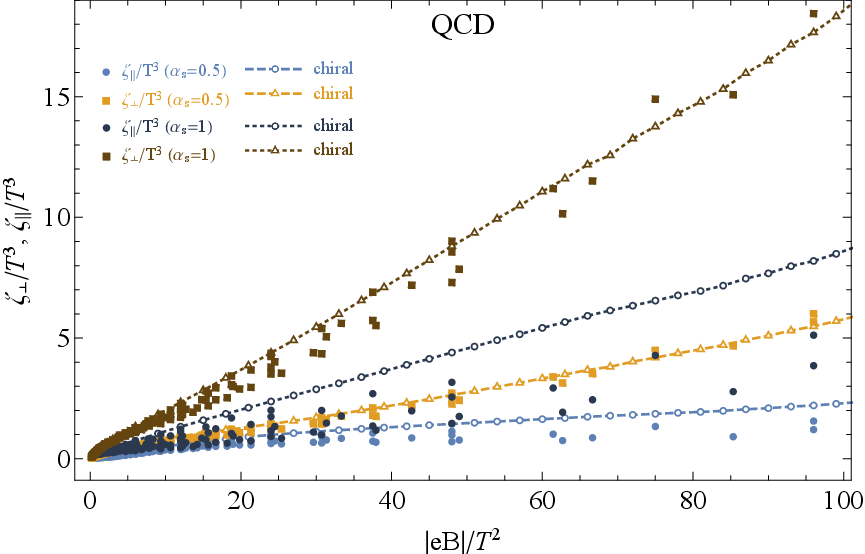} 
\caption{The transverse and longitudinal components of the shear (left) and bulk (right) viscosities as functions of the dimensionless ratio $|eB|/T^2$ in a magnetized QCD plasma for two representative values of the strong coupling constant, $\alpha_s = 0.5$ and $\alpha_s = 1$. Empty markers with interpolating lines correspond to results in the chiral limit, while filled markers represent the massive case.}
\label{fig:viscosity-QCD}
\end{figure}

The numerical results for the transverse and longitudinal components of the shear viscosity in QCD are shown in the left panel of Fig.~\ref{fig:viscosity-QCD}, while the corresponding bulk-viscosity results appear in the right panel. Because we are considering a strongly coupled regime of QCD, the coupling constant is expected to be of order unity, although its precise value is not well constrained. To explore a representative range, we present results for two choices, $\alpha_s =0.5$ (light-colored markers) and $\alpha_s=1$ (dark-colored markers).
The plots contain numerous data points covering a broad span of temperatures, $15 m_q \lesssim T \lesssim 85 m_q$, and magnetic field strengths, $(15 m_q)^{2} \lesssim |eB| \lesssim (225 m_q)^{2}$. Strictly speaking, assuming a deconfined quark-gluon plasma requires $T \gtrsim T_c \simeq 30 m_q $. However, since we present an extrapolation from a weakly-coupled regime of QCD, which cannot capture the nonperturbative dynamics responsible for the deconfinement transition, it is formally acceptable to consider temperatures even below $T_c$, especially in the ultrarelativistic regime where the scaling behavior of the transport coefficients is expected.

Our analysis will focus primarily on the ultra-relativistic regime, $m\ll T$ and $m\ll\sqrt{|eB|}$, which is more amenable to a systematic exploration over a broad parameter space. Specifically, we will consider temperatures and magnetic fields in the ranges $15 m_q\lesssim T\lesssim 85m_q$ and $(15m_q)^{2}\lesssim |eB|\lesssim(225m_q)^{2}$, respectively. We will also demonstrate that, in this regime, the viscosities, when expressed in units of $T^{3}$, obey approximate scaling relations: they are determined by dimensionless functions that depend solely on the ratio $|eB|/T^{2}$. Although these scaling behaviors become exact only in the chiral limit, they remain good approximations throughout the ultra-relativistic domain even for finite fermion mass.

Compared with the QED results in Fig.~\ref{fig:viscosity-QED}, the QCD viscosities exhibit somewhat different behavior, although several qualitative features persist. The transverse shear viscosity remains suppressed relative to its longitudinal counterpart, but the degree of suppression is noticeably weaker than in QED. As a function of the coupling constant, we find that $\eta_{\perp}$ tends to increase, whereas $\eta_{\parallel}$ gets suppressed as $\alpha_s$ grows. Both components of the bulk viscosity remain relative small but appear to continue increasing at large $B$, i.e., they do not decrease as in the QED case. This difference may be the artifact of the strongly coupled nature of QCD, where the effective dimensional reduction induced by the magnetic field is more difficult to realize.

The numerical results for the QCD cross viscosity are shown in the right panel of Fig.~\ref{fig:cross-viscosity-log}. Its qualitative dependence on the dimensionless ratio $|eB|/T^2$ closely resembles the behavior found in QED. It remains negative for all values of the magnetic field and $|\zeta_{\times}|$ is predominantly an increasing function of $|eB|/T^2$. However, its absolute value is generally smaller in strongly coupled QCD. Consistent with this dependence on the coupling strength, the value of $|\zeta_{\times}|$ decreases as $\alpha_s$ increases.

\section{Discussion and Summary}
\label{sec:summary}

In this work, we carried out a detailed quantum field-theoretical study of viscous transport in strongly magnetized relativistic plasmas. By combining Kubo’s linear response formalism with the Landau-level representation of fermionic propagators, and by incorporating the leading-order Landau-level-resolved damping rates, we derived explicit analytical expressions for the transverse, longitudinal, and cross components of both shear and bulk viscosities. These results provide a unified microscopic framework for analyzing magnetoviscous effects in QED and QCD plasmas.

Our numerical investigation of a hot magnetized QED plasma demonstrates that strong magnetic fields induce pronounced anisotropies in the viscous response. The dimensionless longitudinal shear viscosity $\eta_{\parallel}/T^3$  grows rapidly with field strength, while the transverse component $\eta_{\perp}/T^3$  is strongly suppressed, reflecting the very different roles of the lowest Landau level in these channels. The bulk viscosities exhibit nonmonotonic behavior. While they differ, both $\zeta_{\perp}/T^3$ and $\zeta_{\parallel}/T^3$ increase at intermediate fields and decrease toward zero at large fields. Their extreme sensitivity to the longitudinal and transverse components of the sound velocity suggests that quantitative predictions for the bulk viscosity should be interpreted with caution, particularly in the strong-field limit. We also find that the cross viscosity $\zeta_{\times}$ is negative and increases in magnitude with $|eB|$, consistent with expectations from magnetohydrodynamics.

Interestingly, we find that the ratio of transverse shear viscosity to entropy density, $\eta_{\perp}/s$, can fall below the celebrated KSS bound for sufficiently strong magnetic fields, $|eB|/T^{2} \gtrsim 40$. In contrast to previously known examples of anisotropic theories, this violation of the KSS bound occurs in a {\em weakly-coupled} theory. While this may appear surprising, or even seemingly incompatible with the uncertainty principle, we argue that it is a natural consequence of the modified transport mechanism at work in directions transverse to a strong magnetic field. In this regime, transverse momentum transport is mediated by hopping between Landau levels, whose probability is suppressed at weak coupling. In other words, $\eta_{\perp}$ is small because the relevant damping rate, which scales with the coupling strength, is small rather than large.

Extending the analysis to a strongly magnetized quark-gluon plasma, we find that many qualitative trends persist, although quantitative uncertainties grow because the relevant coupling strength is of order unity. The shear viscosity remains highly anisotropic, but the suppression of the transverse component is less dramatic than in QED. The bulk viscosities remain small compared to the shear viscosities, yet, unlike in QED, they do not show a clear tendency to decrease at large magnetic fields. This distinction likely reflects the difficulty of achieving a high-degree dimensional reduction in a strongly coupled QCD plasma. The magnitude of the QCD cross viscosity is also smaller and decreases with increasing $\alpha_s$.

Taken together, these results reveal the rich and highly anisotropic viscous structure of relativistic plasmas in strong magnetic fields. They underscore the importance of including magnetoviscous effects in the interpretation of heavy-ion collision phenomena and in astrophysical applications such as magnetar dynamics. Our findings also provide a benchmark for future studies, including those aimed at incorporating higher-order corrections, improving the treatment of strong coupling, and developing consistent magnetohydrodynamic simulations that include the full tensorial structure of viscosities in magnetized relativistic matter.

The pronounced anisotropy of the shear viscosity suggests that it may substantially influence the observed flow patterns in heavy-ion collisions, thereby reinforcing the earlier conjecture of Ref.~\cite{Tuchin:2011jw}, which was based on a kinetic-theory analysis not applicable in the strong magnetic-field regime. Our results show that the longitudinal shear viscosity can be significantly larger than its transverse counterpart. In the context of heavy-ion physics, this anisotropy can dramatically modify the apparent hydrodynamic flow observables in non-central collisions. A large $\eta_{\parallel}$ is expected to strongly suppress longitudinal shear stresses, whereas a small $\eta_{\perp}$ cannot efficiently dissipate transverse shear stresses. This may, in turn, lead to an enhanced elliptic flow, potentially challenging the conventional assumptions underlying hydrodynamic modeling.

The strong anisotropy of the shear viscosity may also have implications for magnetar magnetospheres. In a first, naive picture, it tends to smooth velocity gradients along magnetic field lines while allowing strong shear across them. Such anisotropic dissipation could modify the dynamics of magnetic flux tubes, current systems, and other magnetospheric structures, and deserves a separate dedicated study.
The viscosity mechanism considered in this work arises from one-to-two and two-to-one processes involving photon emission or absorption. It therefore differs qualitatively from conventional two-to-two scattering, which is expected to play a minor role under magnetospheric conditions. It would be interesting to investigate whether this microscopic mechanism can generate macroscopic effects that are analogous or complementary to radiative drag \cite{Beloborodov:2012sp}.

The strong anisotropy of the shear viscosity may also have implications for magnetar magnetospheres. Within a simplified hydrodynamic picture, and in regimes where such a description is applicable, it may smooth velocity gradients along magnetic field lines more efficiently than across them. This kind of anisotropic dissipation could, in principle, influence the dynamics of magnetic flux tubes, current systems, and related magnetospheric structures, but a detailed study would be required to assess the relevance of these effects. The viscosity mechanism considered in this work arises from $1\leftrightarrow 2$ processes involving photon emission or absorption. In this sense, it differs in its microscopic origin from conventional $2\leftrightarrow 2$  scattering, which we may be subdominant under typical magnetospheric conditions. It would be interesting to explore whether this mechanism can produce macroscopic effects that are analogous or complementary to radiative drag \cite{Beloborodov:2012sp}.

The present study establishes a comprehensive microscopic foundation for understanding viscous transport in strongly magnetized relativistic plasmas, but it also opens several promising avenues for future work. A natural next step is to implement the full anisotropic viscosity tensor into realistic relativistic magnetohydrodynamic simulations of heavy-ion collisions. Such simulations would allow one to quantify how the pronounced splitting between longitudinal and transverse shear viscosities, as well as the nontrivial behavior of the bulk and cross viscosities, influence the evolution of the quark-gluon plasma and the resulting flow observables. In particular, it will be important to assess whether the strong suppression of transverse shear viscosity or the negative cross-bulk viscosity can leave identifiable imprints on charge-dependent flow coefficients, photon and dilepton emissions, or other electromagnetic signatures associated with early-time magnetic fields.
 
\begin{acknowledgments}
This work was supported in part by the U.S. National Science Foundation under Grant Nos.~PHY-2209470 and PHY-2514933. RG is partly supported by Academia Sinica under Project No. AS-CDA-114-M01.
\end{acknowledgments}

\appendix

\section{Thermodynamics expressions for pressure, magnetization, and the speed of sound}
\label{app:Thermodynamics}
 
As seen from Eqs.~(\ref{P-parallel}) through (\ref{v-squared-perp}) in the main text, to define the viscosity coefficients of a magnetized plasma, we require the longitudinal and transverse components of the speed of sound, $v_{s,\parallel}$ and $v_{s,\perp}$. The corresponding expressions are derived in this appendix, together with several related thermodynamic quantities needed in the analysis.

Let us begin with the one-loop (free) pressure of a relativistic plasma in the presence of an external magnetic field $B$, which is given by~\cite{Elmfors:1993wj,Persson:1994pz,Miransky:2015ava}
\begin{equation}
P_{\parallel} = P_g + \sum_{f} \frac{|e_f B| T}{2\pi^2} \sum_{n=0}^\infty \beta_{n} \int_{-\infty}^{\infty}  dp_{z} \ln \left(1 + e^{ - E_{n,f}/T }  \right)  ,
\label{pressure}
\end{equation}
where $ T $ is the temperature, $E_{n,f}=\sqrt{2n|e_{f}B|+p_{z}^2+m_f^2}$ is the Landau-level energy, $ e_f $ and $ m_f $ are the electric charge and mass of the fermion flavor $f$ (which counts all colors and flavors in the case of QCD), and $ \beta_{n} = 2-\delta_{0,n} $. The extra term in the pressure, $P_g$, represents the contribution of gauge bosons, i.e.,
\begin{equation}
P_g = \frac{\pi^2}{45} N_g T^4 ,  
\label{gauge-boson-pressure}
\end{equation}
where $N_g =1$ in QED and $N_g =N_c^2 - 1$ in QCD plasma, respectively.

By making use of the results for the pressure, we derive the following results for the energy density:
\begin{equation}
 \epsilon=T\frac{\partial P}{\partial T}-P
 =3P_g + \sum_{f} \frac{|e_f B| }{2\pi^2} \sum_{n=0}^\infty \beta_{n} \int_{-\infty}^{\infty}  dp_{z} E_{n,f} n_F\left(E_{n,f}\right) .
\end{equation}
Following the same assumption as in the main text, i.e., that the plasma compressions or expansions preserve a magnetic flux \cite{Bali:2014kia}, we have the following  definition of the transverse pressure: $P_{\perp} =P_{\parallel} -M B$, which includes the contribution of the magnetization $ M \equiv \left( \frac{\partial P}{\partial B} \right)_{T} $. By making use of Eq.~(\ref{pressure}), we derive the following expression for the magnetization:
\begin{equation}
M=\left( \frac{\partial P}{\partial B} \right)_{T,\mu} =  q_f  \sum_{f} \frac{T}{2\pi^2} \sum_{n=0}^\infty \beta_{n} \int_{-\infty}^{\infty}  dp_{z}  \left[ \ln \left(1 + e^{ - E_{n,f}/T }  \right) -\frac{n |e_f B|}{TE_{n,f}} n_F\left(E_{n,f}\right)  \right] .
 \label{app:magnetization}
 \end{equation}
Then, the expression for the transverse pressure reads
\begin{equation}
 P_{\perp}=P_g + \sum_{f} \frac{|e_f B|^2 }{2\pi^2} \sum_{n=0}^\infty \beta_{n} \int_{-\infty}^{\infty}  dp_{z} \frac{n_F\left(E_{n,f}\right) }{E_{n,f}} .
\end{equation}
Using the definitions for the longitudinal and transverse components of the speed of sound in Eqs.~(\ref{v-squared-parallel}) and (\ref{v-squared-perp}), we derive the following relations: 
\begin{eqnarray}
v_{s,\perp}^2 &\equiv& v_{s,\parallel}^2 -B \left( \frac{\partial M}{\partial \epsilon} \right)_B
= \frac{s-B\left( \frac{\partial M}{\partial T} \right)_B}{T \left( \frac{\partial s}{\partial T} \right)_B},
\label{v-perp} \\
v_{s,\parallel}^2 &\equiv& \left( \frac{\partial P_{\parallel}}{\partial \epsilon} \right)_B
 =\frac{\left( \frac{\partial P_{\parallel}}{\partial T} \right)_B}{ \left( \frac{\partial \epsilon}{\partial T} \right)_B} 
 =\frac{s}{T \left( \frac{\partial s}{\partial T} \right)_B} .
\label{v-parallel}
\end{eqnarray}
As seen, the results are expressed in terms of the entropy density, as well as its derivative and the derivative of magnetization with respect to temperature, i.e.,  
\begin{eqnarray}
 s&=&\left(\frac{\partial P_{\parallel}}{\partial T}\right)_B
 =\frac{4}{T}P_g+ \frac{1}{2\pi^2}\sum_{f} |e_f B| \sum_{n=0}^\infty \beta_{n} \int_{-\infty}^\infty dp_{z} \left[ \ln \left(1 + e^{ - E_{n,f}/T }  \right)+  \frac{E_{n,f}}{T}n_F\left(E_{n,f}\right)  \right],
 \label{app:entropy} \\
T \left( \frac{\partial s}{\partial T} \right)_B &=& \left( \frac{\partial \epsilon}{\partial T} \right)_B
=\frac{12}{T}P_g+  \frac{1}{8 \pi^2 T^2 }\sum_{f} |e_f B| \sum_{n=0}^\infty \beta_{n} \int_{-\infty}^\infty dp_{z}   \frac{E_{n,f}^2}{\cosh^2\frac{E_{n,f}}{2T}}  ,\\
B \left( \frac{\partial M}{\partial T} \right)_B &=&\frac{1}{2\pi^2}   \sum_{f} |e_f B|  \sum_{n=0}^\infty \beta_{n} \int_{-\infty}^{\infty} dp_{z} \left[ \ln \left(1 + e^{ - E_{n,f}/T }  \right) +\frac{E_{n,f}}{T}n_F\left(E_{n,f}\right) -\frac{n |e_f B|}{4T^2 \cosh^2\frac{E_{n,f}}{2T}}  \right] ,  \\
s-B \left( \frac{\partial M}{\partial T} \right)_B &=& \frac{4}{T}P_g+ \frac{1}{2\pi^2}\sum_{f} |e_f B| \sum_{n=0}^\infty \beta_{n} \int_{-\infty}^\infty dp_{z}   \frac{n |e_f B|}{4T^2 \cosh^2\frac{E_{n,f}}{2T}}  .
\end{eqnarray}

\section{Dirac traces}
\label{app:traces}

In deriving the analytic expressions for the viscosity coefficients, several types of Dirac traces are encountered. For completeness, all such traces are listed here. The first set consists of Dirac traces involving the longitudinal components of the fermion propagators, i.e.,
 \begin{eqnarray}
 T_{aa}^{00} &=& \mbox{tr} \big[ 
 \gamma^0 \left(E_{n} \gamma^{0} -\lambda  k_{z}\gamma^3+\lambda  m  \right)
 \left({\cal P}_{+}L_{n} -{\cal P}_{-}L_{n-1} \right) \gamma^0 \left(E_{n^\prime} \gamma^{0} -\lambda^{\prime}  k_{z}\gamma^3+\lambda^{\prime}  m  \right)
 \left({\cal P}_{+}L_{n^\prime} -{\cal P}_{-}L_{n^\prime-1} \right)  \big]
 \nonumber\\
 &=& 2\left( L_{n}  L_{n^\prime}+L_{n^\prime-1}  L_{n-1}\right) \left[E_{n} E_{n^\prime}+\lambda \lambda^{\prime} (k_{z}^2+
 m^{2})\right] \\
 T_{aa}^{33} &=& \mbox{tr} \big[ 
 \gamma^3 \left(E_{n} \gamma^{0} -\lambda  k_{z}\gamma^3+\lambda  m  \right)
 \left({\cal P}_{+}L_{n} -{\cal P}_{-}L_{n-1} \right)   \gamma^3  \left(E_{n^\prime} \gamma^{0} -\lambda^{\prime}  k_{z}\gamma^3+\lambda^{\prime}  m  \right)
 \left({\cal P}_{+}L_{n^\prime} -{\cal P}_{-}L_{n^\prime-1} \right) \big]
 \nonumber\\
 &=& 2 \left(L_{n}L_{n^\prime}+  L_{n-1}L_{n^\prime-1}\right) \left[E_{n}E_{n^\prime}+\lambda \lambda^{\prime} \left(k_{z}^2-
 m^{2}\right) \right], \\
T_{aa}^{03} &=& T_{aa}^{30} =\mbox{tr} \big[ 
 \gamma^0 \left(E_{n} \gamma^{0} -\lambda  k_{z}\gamma^3+\lambda  m  \right)
 \left({\cal P}_{+}L_{n} -{\cal P}_{-}L_{n-1} \right) \gamma^3 \left(E_{n^\prime} \gamma^{0} -\lambda^{\prime}  k_{z}\gamma^3+\lambda^{\prime}  m  \right)
 \left({\cal P}_{+}L_{n^\prime} -{\cal P}_{-}L_{n^\prime-1} \right)  \big]
 \nonumber\\
 &=& 2 k_{z} \left( L_{n}  L_{n^\prime}+L_{n-1} L_{n^\prime-1}  \right) \left[\lambda^{\prime} E_{n} +\lambda E_{n^\prime} \right], \\
 T_{aa}^{11} &=& \mbox{tr} \big[ 
 \gamma^1 \left(E_{n} \gamma^{0} -\lambda  k_{z}\gamma^3+\lambda  m  \right)
 \left({\cal P}_{+}L_{n} -{\cal P}_{-}L_{n-1} \right) \gamma^1 \left(E_{n^\prime} \gamma^{0} -\lambda^{\prime}  k_{z}\gamma^3+\lambda^{\prime}  m  \right)
 \left({\cal P}_{+}L_{n^\prime} -{\cal P}_{-}L_{n^\prime-1} \right)  \big]
 \nonumber\\
 &=& - 2\left( L_{n}  L_{n^\prime-1}+L_{n^\prime}  L_{n-1}\right) \left[E_{n} E_{n^\prime}-\lambda \lambda^{\prime} (k_{z}^2+
 m^{2})\right]  , \\
 T_{aa}^{12} &=&  - T_{aa}^{21} = \mbox{tr} \big[ 
 \gamma^1 \left(E_{n} \gamma^{0} -\lambda  k_{z}\gamma^3+\lambda  m  \right)
 \left({\cal P}_{+}L_{n} -{\cal P}_{-}L_{n-1} \right)  \gamma^2 \left(E_{n^\prime} \gamma^{0} -\lambda^{\prime}  k_{z}\gamma^3+\lambda^{\prime}  m  \right)
 \left({\cal P}_{+}L_{n^\prime} -{\cal P}_{-}L_{n^\prime-1} \right)  \big] \nonumber\\
 &=& 2 i s_{\perp} \left( L_{n}  L_{n^\prime-1}- L_{n^\prime}  L_{n-1}\right) \left[E_{n} E_{n^\prime}-\lambda \lambda^{\prime} 
 (k_{z}^2+m^{2})\right] , \\
 T_{aa}^{\perp,\perp} &=& \mbox{tr} \big[ 
 \frac{\bm{k}_{\perp}\cdot \bm{\gamma}_{\perp}}{2} 
 \left(E_{n} \gamma^{0} -\lambda  k_{z}\gamma^3+\lambda  m  \right)
 \left({\cal P}_{+}L_{n} -{\cal P}_{-}L_{n-1} \right) 
 \frac{\bm{k}_{\perp}\cdot \bm{\gamma}_{\perp}}{2}
 \left(E_{n^\prime} \gamma^{0} -\lambda^{\prime}  k_{z}\gamma^3+\lambda^{\prime}  m  \right)
 \left({\cal P}_{+}L_{n^\prime} -{\cal P}_{-}L_{n^\prime-1} \right)  \big]
 \nonumber\\
 &=&-\frac{k_{\perp}^2}{2}\left( L_{n}  L_{n^\prime-1}+L_{n^\prime}  L_{n-1}\right) \left[E_{n} E_{n^\prime}-\lambda \lambda^{\prime} (k_{z}^2+
 m^{2})\right]  , \\
T_{aa}^{01} &=& T_{aa}^{10}= T_{aa}^{02}=T_{aa}^{20}= T_{aa}^{13}= T_{aa}^{31}= T_{aa}^{23}=T_{aa}^{32}=0 .
 \end{eqnarray}
The second set contains Dirac traces involving the transverse components of the fermion propagators, i.e.,
 \begin{eqnarray}
T_{bb}^{00} &=& 4\lambda \lambda^{\prime} \mbox{tr} \left[ 
 \gamma^0 (\bm{k}_{\perp}\cdot\bm{\gamma}_{\perp}) L_{n-1}^{1}  
 \gamma^0 (\bm{k}_{\perp}\cdot\bm{\gamma}_{\perp}) L_{n^{\prime} -1}^{1}   \right]
 = 16 \lambda \lambda^{\prime} k_{\perp}^2 L_{n-1}^{1}  L_{n^{\prime} -1}^{1}  , \\
T_{bb}^{33}  &=& 4\lambda \lambda^{\prime} \mbox{tr} \left[ 
 \gamma^3 (\bm{k}_{\perp}\cdot\bm{\gamma}_{\perp}) L_{n-1}^{1}  
 \gamma^3 (\bm{k}_{\perp}\cdot\bm{\gamma}_{\perp}) L_{n^{\prime}-1}^{1}   \right]
 =-16 \lambda \lambda^{\prime} k_{\perp}^2  L_{n-1}^{1}  L_{n^{\prime} -1}^{1} , \\
T_{bb}^{11} &=& -  T_{bb}^{22} = 4\lambda \lambda^{\prime} \mbox{tr} \left[ 
 \gamma^1 (\bm{k}_{\perp}\cdot\bm{\gamma}_{\perp}) L_{n-1}^{1}  
 \gamma^1 (\bm{k}_{\perp}\cdot\bm{\gamma}_{\perp}) L_{n^{\prime} -1}^{1}   \right]
 = 16 \lambda \lambda^{\prime} \left(k_{x}^2-k_{y}^2\right) L_{n-1}^{1}  L_{n^{\prime} -1}^{1} 
 \label{Tbb11} , \\
T_{bb}^{12} &=& T_{bb}^{21} = 4\lambda \lambda^{\prime} \mbox{tr} \left[ 
 \gamma^1 (\bm{k}_{\perp}\cdot\bm{\gamma}_{\perp}) L_{n-1}^{1}  
 \gamma^2 (\bm{k}_{\perp}\cdot\bm{\gamma}_{\perp}) L_{n^{\prime} -1}^{1}   \right]
 =32 \lambda \lambda^{\prime}  k_{x} k_{y} L_{n-1}^{1}  L_{n^{\prime} -1}^{1}  , \\
T_{bb}^{\perp,\perp} &=& 4\lambda \lambda^{\prime} \mbox{tr} \left[ 
 \frac{\bm{k}_{\perp}\cdot \bm{\gamma}_{\perp}}{2} 
 (\bm{k}_{\perp}\cdot\bm{\gamma}_{\perp}) L_{n-1}^{1}  
 \frac{\bm{k}_{\perp}\cdot \bm{\gamma}_{\perp}}{2} 
  (\bm{k}_{\perp}\cdot\bm{\gamma}_{\perp}) L_{n^{\prime} -1}^{1}   \right]
 = 4 \lambda \lambda^{\prime} k_{\perp}^4 L_{n-1}^{1}  L_{n^{\prime} -1}^{1}  , \\
T_{bb}^{03}&=& T_{bb}^{30}=T_{bb}^{10}=T_{bb}^{01} =T_{bb}^{20}=T_{bb}^{02}
 =T_{bb}^{13}=T_{bb}^{31} =T_{bb}^{23}=T_{bb}^{32}=0 .
 \end{eqnarray}
Finally, the last two sets correspond to the mixed Dirac traces involving both longitudinal and transverse components of the fermion propagators, i.e., 
\begin{eqnarray}
 T_{ab}^{01} &=&  2\lambda^{\prime}  \mbox{tr} \left[ 
 \gamma^0 \left(E_{n} \gamma^{0} -\lambda  k_{z}\gamma^3+\lambda  m  \right)
 \left({\cal P}_{+}L_{n} -{\cal P}_{-}L_{n-1} \right) 
 \gamma^1 (\bm{k}_{\perp}\cdot\bm{\gamma}_{\perp}) L_{n^{\prime} -1}^{1}   \right] \nonumber\\
 &=& -4 \lambda^{\prime} E_{n}  
 \left[ k_{x}\left(L_{n} - L_{n-1} \right) + i s_{\perp} k_{y} \left(L_{n} + L_{n-1} \right) \right] L_{n^{\prime} -1}^{1} ,\\ 
 T_{ab}^{10} &=&  2\lambda^{\prime}  \mbox{tr} \left[ 
 \gamma^1 \left(E_{n} \gamma^{0} -\lambda  k_{z}\gamma^3+\lambda  m  \right)
 \left({\cal P}_{+}L_{n} -{\cal P}_{-}L_{n-1} \right) 
 \gamma^0 (\bm{k}_{\perp}\cdot\bm{\gamma}_{\perp}) L_{n^{\prime} -1}^{1}   \right] \nonumber\\
 &=& -4 \lambda^{\prime} E_{n}  
 \left[ k_{x}\left(L_{n} - L_{n-1} \right) - i s_{\perp} k_{y} \left(L_{n} + L_{n-1} \right) \right] L_{n^{\prime} -1}^{1} ,\\
  T_{ab}^{13} &=&  2\lambda^{\prime}  \mbox{tr} \left[ 
 \gamma^1 \left(E_{n} \gamma^{0} -\lambda  k_{z}\gamma^3+\lambda  m  \right)
 \left({\cal P}_{+}L_{n} -{\cal P}_{-}L_{n-1} \right) 
 \gamma^3 (\bm{k}_{\perp}\cdot\bm{\gamma}_{\perp}) L_{n^{\prime} -1}^{1}   \right] \nonumber\\
 &=& -4 \lambda\lambda^{\prime} k_{z}  \left[ k_{x}\left(L_{n} - L_{n-1} \right) -i s_{\perp} k_{y} \left(L_{n} + L_{n-1} \right) \right] L_{n^{\prime} -1}^{1} ,\\
 T_{ab}^{31} &=&  2\lambda^{\prime}  \mbox{tr} \left[ 
 \gamma^3 \left(E_{n} \gamma^{0} -\lambda  k_{z}\gamma^3+\lambda  m  \right)
 \left({\cal P}_{+}L_{n} -{\cal P}_{-}L_{n-1} \right) 
 \gamma^1 (\bm{k}_{\perp}\cdot\bm{\gamma}_{\perp}) L_{n^{\prime} -1}^{1}   \right] \nonumber\\
 &=& -4 \lambda\lambda^{\prime} k_{z}  \left[ k_{x}\left(L_{n} - L_{n-1} \right) + i s_{\perp} k_{y} \left(L_{n} + L_{n-1} \right) \right] L_{n^{\prime} -1}^{1} ,\\
T_{ab}^{00} &=&T_{ab}^{33} =T_{ab}^{03}= T_{ab}^{30} = T_{ab}^{11} =T_{ab}^{22}  =T_{ab}^{12} =T_{ab}^{21}=0 ,
 \end{eqnarray}
and
  \begin{eqnarray}
  T_{ba}^{01} &=&  2\lambda  \mbox{tr} \left[ 
 \gamma^0 (\bm{k}_{\perp}\cdot\bm{\gamma}_{\perp}) L_{n-1}^{1}  
 \gamma^1 \left(E_{n^\prime} \gamma^{0} -\lambda^{\prime}  k_{z}\gamma^3+\lambda^{\prime}  m  \right)
 \left({\cal P}_{+}L_{n^\prime} -{\cal P}_{-}L_{n^\prime-1} \right)  \right] \nonumber\\
 &=& -4 \lambda E_{n^{\prime}}  
 \left[ k_{x}\left(L_{n^\prime} - L_{n^\prime-1} \right) - i s_{\perp} k_{y} \left(L_{n^\prime} + L_{n^\prime-1} \right) \right] L_{n-1}^{1} ,\\ 
 T_{ba}^{10} &=&  2\lambda  \mbox{tr} \left[ 
 \gamma^1 (\bm{k}_{\perp}\cdot\bm{\gamma}_{\perp}) L_{n-1}^{1}  
 \gamma^0 \left(E_{n^\prime} \gamma^{0} -\lambda^{\prime}  k_{z}\gamma^3+\lambda^{\prime}  m  \right)
 \left({\cal P}_{+}L_{n^\prime} -{\cal P}_{-}L_{n^\prime-1} \right)  \right] \nonumber\\
 &=& -4 \lambda E_{n^{\prime}}  
 \left[ k_{x}\left(L_{n^\prime} - L_{n^\prime-1} \right) + i s_{\perp} k_{y} \left(L_{n^\prime} + L_{n^\prime-1} \right) \right] L_{n -1}^{1} ,\\
 T_{ba}^{13} &=&  2\lambda  \mbox{tr} \left[ 
 \gamma^1 (\bm{k}_{\perp}\cdot\bm{\gamma}_{\perp}) L_{n-1}^{1}  
 \gamma^3 \left(E_{n^\prime} \gamma^{0} -\lambda^{\prime}  k_{z}\gamma^3+\lambda^{\prime}  m  \right)
 \left({\cal P}_{+}L_{n^\prime} -{\cal P}_{-}L_{n^\prime-1} \right)  \right] \nonumber\\
 &=& -4 \lambda\lambda^{\prime} k_{z}  \left[ k_{x}\left(L_{n^\prime} - L_{n^\prime-1} \right) +i s_{\perp} k_{y} \left(L_{n^\prime} + L_{n^\prime-1} \right) \right] L_{n-1}^{1} ,\\
 T_{ba}^{31} &=&  2\lambda  \mbox{tr} \left[ 
 \gamma^3 (\bm{k}_{\perp}\cdot\bm{\gamma}_{\perp}) L_{n-1}^{1}  
 \gamma^1 \left(E_{n^\prime} \gamma^{0} -\lambda^{\prime}  k_{z}\gamma^3+\lambda^{\prime}  m  \right)
 \left({\cal P}_{+}L_{n^\prime} -{\cal P}_{-}L_{n^\prime-1} \right)  \right] \nonumber\\
 &=& -4 \lambda\lambda^{\prime} k_{z}  \left[ k_{x}\left(L_{n^\prime} - L_{n^\prime-1} \right) - i s_{\perp} k_{y} \left(L_{n^\prime} + L_{n^\prime-1} \right) \right] L_{n-1}^{1} ,\\
T_{ba}^{00} &=&T_{ba}^{33} =T_{ba}^{03}= T_{ba}^{30} = T_{ba}^{11} =T_{ba}^{22}  =T_{ba}^{12} =T_{ba}^{21}=0 ,
 \end{eqnarray}
respectively.

\section{Analytic results for transverse momentum integrals}
\label{app:integrals}

In performing the integration over the transverse momentum $k_{\perp}$ appearing in Kubo's expressions for the viscosity coefficients, we used the following table integrals:
 \begin{eqnarray}
K_{1} &=&\int_{0}^{\infty}   k_{\perp} dk_{\perp} e^{-2k_{\perp}^2\ell^2} L_{n}\left(2 k_{\perp}^2 \ell^2\right) L_{n^\prime} \left(2 k_{\perp}^2 \ell^2\right) 
 = \frac{1}{4\ell^2} \int_{0}^{\infty} dx e^{-x}L_{n}(x) L_{n^\prime}(x) = \frac{1}{4\ell^2} \delta_{n,n^\prime} , \\
K_{2} &=&\int_{0}^{\infty}  k_{\perp}^3 dk_{\perp} e^{-2k_{\perp}^2\ell^2} L_{n}\left(2 k_{\perp}^2 \ell^2\right) L_{n^\prime} \left(2 k_{\perp}^2 \ell^2\right)
 = \frac{1}{8\ell^4} \int_{0}^{\infty} x dx e^{-x}L_{n}(x) L_{n^\prime}(x) \nonumber\\
 &=&  \frac{1}{8\ell^4} 
 \left[(n+n^\prime+1)\delta_{n,n^\prime} - n^\prime \delta_{n+1,n^\prime} -n \delta_{n,n^\prime+1}  \right], \\
K_{3} &=&\int_{0}^{\infty}  k_{\perp}^3 dk_{\perp} e^{-2k_{\perp}^2\ell^2} L_{n} \left(2 k_{\perp}^2 \ell^2\right) L_{n^{\prime}-1 }^{1} \left(2 k_{\perp}^2 \ell^2\right)
 =  - \frac{n^\prime }{8\ell^4} \left(  \delta_{n,n^\prime} -\delta_{n+1,n^\prime} \right) , \\
K_4 &=&\int_{0}^{\infty}  k_{\perp}^3 dk_{\perp} e^{-2k_{\perp}^2\ell^2} L_{n-1}^{1}  \left(2 k_{\perp}^2 \ell^2\right) L_{n^{\prime} -1}^{1} \left(2 k_{\perp}^2 \ell^2\right)
 =\frac{1}{8\ell^4} \int_{0}^{\infty} x dx e^{-x}L_{n-1}^{1} (x) L_{n^\prime-1}^{1} (x) =  \frac{n}{8\ell^4} \delta_{n,n^\prime} , \\
K_5 &=& \int_{0}^{\infty}   k_{\perp}^5 dk_{\perp} e^{-2k_{\perp}^2\ell^2} L_{n-1}^{1}  \left(2 k_{\perp}^2 \ell^2\right) L_{n^{\prime} -1}^{1} \left(2 k_{\perp}^2 \ell^2\right)
 =\frac{1}{16\ell_{f}^6} \int_{0}^{\infty} x^2 dx e^{-x}L_{n-1}^{1} (x) L_{n^\prime-1}^{1} (x)  \nonumber\\
 &=&  \frac{n n^\prime }{16\ell_{f}^6} \left( 2 \delta_{n,n^\prime} - \delta_{n+1,n^\prime} - \delta_{n,n^\prime+1}  \right)  .
\end{eqnarray}
It is worth noting that similar integrals were presented in Ref.~\cite{Ghosh:2020wqx}. However, we found that several of the reported results contained errors. 

By combining the integrals listed above, we can further obtain the following results:
\begin{eqnarray}
K_6 &=& \int_{0}^{\infty} k_{\perp}^3 dk_{\perp} e^{-2k_{\perp}^2\ell^2} \left( L_{n}  L_{n^\prime-1}+L_{n^\prime}  L_{n-1}\right)
=  -\frac{n+n^\prime}{8\ell^4}\left(
\delta_{n,n^\prime}-\delta_{n+1,n^\prime}-\delta_{n,n^\prime+1}+\frac{1}{2}\delta_{n+2,n^\prime}+\frac{1}{2}\delta_{n,n^\prime+2}
\right) ,\\
K_7 &=& \int_{0}^{\infty} k_{\perp}^3 dk_{\perp} e^{-2k_{\perp}^2\ell^2} \left(L_{n} - L_{n-1} \right) L^{1}_{n^{\prime}-1} 
=  - \frac{n^{\prime}}{8\ell^4}\left(2\delta_{n,n^\prime}-\delta_{n+1,n^\prime}-\delta_{n,n^\prime+1} \right)  .
\end{eqnarray}

\section{Evaluation of the $\lambda$ and $\lambda^\prime$ sums}
\label{app:sums}

In deriving the viscosity coefficients, the final expressions involve summations over the energy-sign indices, $\lambda$ and $\lambda^{\prime}$. In total, four distinct types of sums appear. For convenience, their results are listed here:
\begin{eqnarray}
\sum_{\lambda,\lambda^\prime=\pm1}  \rho(k_{0}, \lambda E_{n})\rho(k_{0}, \lambda^\prime E_{n^\prime}) 
 &=& \frac{4 \Gamma_{n} \Gamma_{n^\prime} }{\pi^2}\frac{\left(k_{0}^2+E_{n}^2+\Gamma_{n}^2\right)\left(k_{0}^2+E_{n^\prime}^2+\Gamma_{n^\prime}^2\right) }
 {\left[\left(E_{n}^2+\Gamma_{n}^2-k_{0}^2\right)^2+4 k_{0}^2\Gamma_{n}^2\right]\left[\left(E_{n^\prime}^2+\Gamma_{n^\prime}^2-k_{0}^2\right)^2+4 k_{0}^2\Gamma_{n^\prime}^2\right]},
\end{eqnarray}
\begin{eqnarray}
\sum_{\lambda,\lambda^\prime=\pm1} \lambda \lambda^{\prime}  \rho(k_{0}, \lambda E_{n})\rho(k_{0}, \lambda^\prime E_{n^\prime}) 
 &=&  \frac{4 \Gamma_{n} \Gamma_{n^\prime} }{\pi^2}\frac{ 4k_{0}^2 E_{n}E_{n^\prime} }
 {\left[\left(E_{n}^2+\Gamma_{n}^2-k_{0}^2\right)^2+4 k_{0}^2\Gamma_{n}^2\right]\left[\left(E_{n^\prime}^2+\Gamma_{n^\prime}^2-k_{0}^2\right)^2+4 k_{0}^2\Gamma_{n^\prime}^2\right]},
\end{eqnarray}
\begin{eqnarray}
\sum_{\lambda,\lambda^\prime=\pm1} \frac{\lambda^\prime k_{0}}{E_{n^\prime} }   \rho(k_{0}, \lambda E_{n})\rho(k_{0}, \lambda^\prime E_{n^\prime}) 
 &=& \frac{4 \Gamma_{n} \Gamma_{n^\prime} }{\pi^2} \frac{2 k_{0}^2 \left(E_{n}^2+\Gamma_{n}^2+k_{0}^2\right) }
 {\left[\left(E_{n}^2+\Gamma_{n}^2-k_{0}^2\right)^2+4 k_{0}^2\Gamma_{n}^2\right]\left[\left(E_{n^\prime}^2+\Gamma_{n^\prime}^2-k_{0}^2\right)^2+4 k_{0}^2\Gamma_{n^\prime}^2\right]},
\end{eqnarray}
\begin{eqnarray}
\sum_{\lambda,\lambda^\prime=\pm1}  \frac{\lambda k_{0}}{E_{n} }  \rho(k_{0}, \lambda E_{n})\rho(k_{0}, \lambda^\prime E_{n^\prime}) 
 &=& \frac{4 \Gamma_{n} \Gamma_{n^\prime} }{\pi^2} \frac{2 k_{0}^2 \left(E_{n^\prime}^2+\Gamma_{n^\prime}^2+k_{0}^2\right)}
 {\left[\left(E_{n}^2+\Gamma_{n}^2-k_{0}^2\right)^2+4 k_{0}^2\Gamma_{n}^2\right]\left[\left(E_{n^\prime}^2+\Gamma_{n^\prime}^2-k_{0}^2\right)^2+4 k_{0}^2\Gamma_{n^\prime}^2\right]}.
\end{eqnarray}

\section{Transverse and longitudinal viscosities in the limit $B\to 0$}
\label{app:limit-B0}

To cross-check the definitions of the viscosities in a magnetized plasma, it is instructive to verify that the transverse and longitudinal components reduce to identical forms in the limit $B \to 0$. For simplicity, we consider here the contribution from a single flavor of charged fermions. In the presence of multiple flavors (and/or colors), the corresponding degeneracy factors should be included accordingly.

\subsection{Shear viscosities in the limit $B\to 0$}
\label{app:limit-B0-shear}

Let us begin with the components of the shear viscosity given in Eqs.~(\ref{shear-perp-1}) and (\ref{shear-parallel-1}) of the main text. In the limit of a vanishing magnetic field, the discrete sum over the Landau-level index $n$ can be replaced by an integral over the continuous variable $x = 2n|e_f B|$, which corresponds to the squared transverse momentum in the $B=0$ case. Specifically, by employing the summation formula
\begin{equation}
\sum_{n=0}^{\infty} f(2n |e_{f}B|) = \frac{\ell^2}{2} \int_{0}^{\infty} dx f(x)
= \ell^2 \int_{0}^{\infty} k_{\perp} dk_{\perp} f(k_{\perp}^2)
= \ell^2 \int \frac{d^2\bm{k}_{\perp}}{2\pi} f(k_{\perp}^2) ,
\label{sumLL-integral}
\end{equation}
and substituting it into the expressions for the shear viscosities, we derive
\begin{equation}
 \eta_{\perp} = \frac{1}{128 \pi^2 T} \sum_{\lambda,\lambda^\prime=\pm} 
 \int \frac{d^3\bm{k} dk_{0}}{\cosh^2\frac{k_{0}}{2T}} 
 \rho(k_{0}, \lambda E_{k})\rho(k_{0}, \lambda^\prime E_{k}) 
 k_{\perp}^2
 \left(1-\lambda \lambda^{\prime} \frac{k_{z}^2+ m^{2}}{E_{k}^2 }\right)  ,
 \label{shear-perp-B0a}
\end{equation}
for the transverse component, and
\begin{equation}
 \eta_{\parallel} = \frac{1}{256 \pi^2 T}  
 \sum_{\lambda,\lambda^\prime=\pm} 
 \int \frac{d^3\bm{k} dk_{0}}{\cosh^2\frac{k_{0}}{2T}}  
 \rho(k_{0}, \lambda E_{k})\rho(k_{0}, \lambda^\prime E_{k})
 \Bigg\{k_{\perp}^2
 \left(1+\lambda \lambda^{\prime} \frac{k_{z}^2- m^{2}}{E_{k}^2} \right) 
  - \frac{k_{\perp}^4 \lambda \lambda^{\prime}}{E_{k}^2}   
 +2k_{z}^2  \left(1-\lambda \lambda^{\prime} \frac{k_{z}^2+ m^{2}}{E_{k}^2}\right) 
 + \frac{4 k_{\perp}^2 \lambda\lambda^{\prime} k_{z}^2}{E_{k}^2}   \Bigg\}, 
\label{shear-parallel-B0a}
\end{equation}
for the longitudinal component, respectively. Furthermore, substituting $m^2 = E_{k}^2 - k^2$ and averaging over the angular coordinates, we obtain
\begin{equation}
 \eta_{\perp} = \frac{1}{192 \pi^2 T} \sum_{\lambda,\lambda^\prime=\pm} 
 \int \frac{d^3\bm{k} dk_{0}}{\cosh^2\frac{k_{0}}{2T}} 
 \rho(k_{0}, \lambda E_{k})\rho(k_{0}, \lambda^\prime E_{k}) 
 \left[ k^2 \left(1-\lambda \lambda^{\prime}\right)  
 +\frac{4}{5}  \lambda \lambda^{\prime} \frac{k^4 }{E_{k}^2} \right], 
 \label{shear-perp-B0b}
\end{equation}
along with an identical result for $\eta_{\parallel}$. Note that averaging over the angular coordinates is equivalent to making the following substitutions: $k_{z}^2\to \frac{1}{3}k^2$, $k_{\perp}^2\to \frac{2}{3}k^2$, $k_{\perp}^2 k_{z}^2\to \frac{2}{15}k^4$, $k_{z}^4 \to \frac{1}{5}k^4$, and $k_{\perp}^4 \to \frac{8}{15}k^4$.

We can further verify the result by applying Kubo's formulas in Eqs.~(\ref{shear-perp-AA}) and (\ref{shear-parallel-AA}) in the absence of a background magnetic field. In this case, the relevant quantities are determined by the spectral function of the fermion propagators at $B=0$,
\begin{eqnarray}
A_{\bm{k}} (k_{0}) &=&  \sum_{\lambda=\pm}  \frac{i}{2 E_{k}}  \left[E_{k} \gamma^{0} 
 -\lambda ( \bm{k}\cdot\bm{\gamma}) +\lambda  m \right]  \rho(k_{0}, \lambda E_{k}) .   
 \label{spectral-density-B0}
\end{eqnarray}
Substituting this spectral function into Eqs.~(\ref{shear-perp-AA}) and (\ref{shear-parallel-AA}), we obtain the following expressions for the transverse and longitudinal shear viscosities:
\begin{eqnarray}
\eta_{\perp} &=&  \frac{1}{64 \pi^2 T} \sum_{\lambda,\lambda^\prime=\pm}
 \int \frac{d^3\bm{k} d k_{0}}{\cosh^2\frac{k_{0}}{2T}}  \rho(k_{0}, \lambda E_{k})\rho(k_{0}, \lambda^{\prime} E_{k}) 
\left[ k_{y}^2 (1-\lambda\lambda^\prime)
+4\lambda\lambda^\prime \frac{k_{x}^2 k_{y}^2}{E_{k}^2}  \right] ,
\end{eqnarray}
and 
\begin{eqnarray}
\eta_{\parallel} &=& \frac{1}{64 \pi^2 T}  \sum_{\lambda,\lambda^\prime=\pm}
 \int \frac{d^3\bm{k} d k_{0}}{\cosh^2\frac{k_{0}}{2T}}  \rho(k_{0}, \lambda E_{k})\rho(k_{0}, \lambda^{\prime} E_{k}) 
\left[  \frac{1}{2}(k_{x}^2+k_{z}^2)  (1-\lambda\lambda^\prime)+4 \lambda\lambda^\prime \frac{k_{x}^2 k_{z}^2}{E_{k}^2} \right].\end{eqnarray}
After averaging over the angular coordinates, both expressions reduce to the same form, as expected, $\eta_{\perp} = \eta_{\parallel} =\eta_{0}$ with
\begin{equation}
\eta_{0} =\frac{1}{192 \pi^2 T}  \sum_{\lambda,\lambda^\prime=\pm}
 \int \frac{d^3\bm{k} d k_{0}}{\cosh^2\frac{k_{0}}{2T}}  \rho(k_{0}, \lambda E_{k})\rho(k_{0}, \lambda^{\prime} E_{k}) 
\left[ k^2 (1-\lambda\lambda^\prime)
+\frac{4}{5}\lambda\lambda^\prime \frac{k^4}{E_{k}^2}
\right].
 \label{shear-perp-B0c}
\end{equation}
As seen, this also agrees with the result in Eq.~(\ref{shear-perp-B0b}) obtained for a magnetized plasma in the limit $B \to 0$.

Let us mention in passing that, in the weak-coupling limit where $\Gamma_{k} \to 0$, one may approximate the spectral function as $\rho(k_{0}, \lambda E_{k}) \simeq \delta(k_{0} - \lambda E_{k})$ and carry out the integration over $k_{0}$ to obtain
\begin{equation}
\eta_{0} \simeq  \frac{1}{30 \pi^2 T} \int \frac{k^6dk}{E_{k}^2 \Gamma_{k} \cosh^2\frac{E_{k}}{2T}}.
\end{equation}
If the $k$-dependence of the damping rate $\Gamma_{k} $ is neglected and the quasiparticle mass is set to zero, this further simplifies to
\begin{equation}
\eta_{0} \simeq \frac{1}{30 \pi^2 T \Gamma} \int \frac{k^4dk}{\cosh^2\frac{k}{2T}} 
=\frac{7\pi^2 T^4}{225 \Gamma} .
\end{equation}

\subsection{Bulk viscosities in the limit $B\to 0$}
\label{app:limit-B0-bulk}

Let us now examine the limiting expressions for the transverse and longitudinal components of the bulk viscosity. Similar to the case of shear viscosity, the sum over Landau levels in Eqs.~(\ref{zeta-perp-00}) and (\ref{zeta-parallel-00}) can be replaced by an integral. Employing the same summation formula, Eq.~(\ref{sumLL-integral}), and substituting $m^2 = E_{k}^2 - k^2$, we obtain
\begin{eqnarray}
 \zeta_{\perp} &=& \frac{1}{192 \pi^2 T} 
 \sum_{\lambda,\lambda^\prime=\pm}\int \frac{dk_{0} d^3\bm{k}}{\cosh^2\frac{k_{0}}{2T}}  
 \rho(k_{0}, \lambda E_{k})\rho(k_{0}, \lambda^\prime E_{k}) \Bigg[
k_{\perp}^2  \left(1-\lambda\lambda^{\prime} +2 \lambda \lambda^{\prime} \frac{k^{2}}{E_{k}^2} 
 \right) \nonumber\\
 &+&6 v_{s}^4  k_{0}^2 \left(1+\lambda \lambda^{\prime}  \right) 
  -2v_{s}^2   k_{z}^2 k_{0} \frac{\lambda+\lambda^{\prime} }{E_{k}} 
  - 5 v_{s}^2 k_{\perp}^2 k_{0} \frac{\lambda^{\prime} +\lambda}{E_{k}}  
  \Bigg],
\end{eqnarray}
and 
\begin{eqnarray}
 \zeta_{\parallel} &=&  \frac{1}{96 \pi^2 T} 
 \sum_{\lambda,\lambda^\prime=\pm} 
 \int  \frac{ dk_{0} d^3 \bm{k}}{\cosh^2\frac{k_{0}}{2T}}  
 \rho(k_{0}, \lambda E_{k}) \rho(k_{0}, \lambda^\prime E_{k})
 \Bigg[k_{z}^2 
 \left(1 -\lambda \lambda^{\prime} +2\lambda \lambda^{\prime} \frac{k^2}{E_{k}^2 }   \right)  
 +3 v_{s}^4  k_{0}^2 
 \left(1+\lambda \lambda^{\prime} \right)  \nonumber\\
 &-&v_{s}^2 (4 k_{z}^2 + k_{\perp}^2) k_{0} \frac{\lambda +\lambda^{\prime}}{E_{k}} 
 \Bigg] ,
\end{eqnarray}
where we have assumed an isotropic speed of sound, i.e., $v_{s,\perp}^2  =v_{s,\parallel}^2  =v_{s}^2$.
Performing the angular average then yields
\begin{equation}
 \zeta_{\perp} = \frac{1}{288  \pi^2 T} 
 \sum_{\lambda,\lambda^\prime=\pm}\int \frac{dk_{0} d^3\bm{k}}{\cosh^2\frac{k_{0}}{2T}}  
 \rho(k_{0}, \lambda E_{k})\rho(k_{0}, \lambda^\prime E_{k}) \Bigg[
 k^2 \left(1- \lambda\lambda^{\prime}\right) 
 +9  v_{s}^4  k_{0}^2 \left(1+\lambda \lambda^{\prime}  \right) 
  -6 v_{s}^2 k^2 k_{0} \frac{\lambda^{\prime} +\lambda}{E_{k}}
 + 2 \lambda\lambda^{\prime} \frac{k^4}{E_{k}^2}   \Bigg],
 \label{bulk-perp-B0b}
\end{equation}
along with an identical result for $\zeta_{\parallel}$.

Using the relations $\lambda \lambda^{\prime}(\lambda + \lambda^{\prime}) = \lambda + \lambda^{\prime}$, $(\lambda + \lambda^{\prime})^2 = 2(1 + \lambda \lambda^{\prime})$, and $\lambda \lambda^{\prime}(1 + \lambda \lambda^{\prime}) = 1 + \lambda \lambda^{\prime}$, one can prove the identity
\begin{equation}
2\lambda \lambda^{\prime} \left(\frac{k^2}{E_{k}} -3 v_{s}^2 k_{0}  \frac{\lambda+ \lambda^{\prime}}{2} \right)^2
=2 \lambda \lambda^{\prime} \frac{k^4}{E_{k}^2} -6v_{s}^2 \frac{k^2}{E_{k}}  k_{0} \left(\lambda+ \lambda^{\prime} \right)
+9 v_{s}^4 k_{0}^2 \left(1+ \lambda \lambda^{\prime} \right) .
\label{kk-square}
\end{equation}
With the help of this relation, the expression in Eq.~(\ref{bulk-perp-B0b}) for the bulk viscosity in the $B \to 0$ limit can be recast in the following positive-definite form:
\begin{equation}
 \zeta_{0} = \frac{1}{288 \pi^2 T} 
 \sum_{\lambda,\lambda^\prime=\pm}\int \frac{dk_{0} d^3\bm{k}}{\cosh^2\frac{k_{0}}{2T}}  
 \rho(k_{0}, \lambda E_{k})\rho(k_{0}, \lambda^\prime E_{k}) \left[
 k^2 \left(1- \lambda\lambda^{\prime}\right) 
 +2\lambda \lambda^{\prime} \left(\frac{k^2}{E_{k}} -3 v_{s}^2 k_{0}  \frac{\lambda+ \lambda^{\prime}}{2} \right)^2   \right].
\label{bulk-perp-B0c}
\end{equation}
Furthermore, we can verify that the same result is obtained from both Kubo relations, Eqs.~(\ref{zeta-perp-0}) and (\ref{zeta-para-0}), when the zero-field spectral function, Eq.~(\ref{spectral-density-B0}), is utilized, i.e., $\zeta_{\perp} = \zeta_{\parallel} =\zeta_{0}$ at $B=0$. 

Interestingly, in the weak-coupling limit where $\Gamma_{k} \to 0$, one may approximate the spectral function as $\rho(k_{0}, E_{k}) \simeq \delta(k_{0} - E_{k})$ and carry out the integration over $k_{0}$ to obtain
\begin{equation}
\zeta_{0} \simeq  \frac{1}{36 \pi^2 T} \int \frac{E_{k}^2 k^2dk}{\Gamma_{k} \cosh^2\frac{E_{k}}{2T}} \left(\frac{k^2}{E_{k}^2} -3 v_{s}^2  \right)^2 .
\label{zeta0}
\end{equation}
If, in addition, the $k$-dependence of the damping rate is neglected and the quasiparticle mass is set to zero, this further simplifies to
\begin{equation}
\zeta_{0} \simeq \frac{1}{36 \pi^2 T \Gamma} \int \frac{k^4dk}{\cosh^2\frac{k}{2T}} \left(1 -3 v_{s}^2  \right)^2 
=\frac{7\pi^2 T^4}{270 \Gamma} \left(1 -3 v_{s}^2  \right)^2 .
\end{equation}

\subsection{Cross viscosity in the limit $B\to 0$}
\label{app:limit-B0-cross}

Let us now examine the limiting form of the cross viscosity. Starting from the analytic expression in Eq.~(\ref{zeta-cross-01}) and replacing the discrete sum over Landau levels with an integral using the summation formula in Eq.~(\ref{sumLL-integral}), we obtain
\begin{equation}
\zeta_{\times}^{(B\to 0)} = \frac{1}{64\pi^2 T } \sum_{\lambda,\lambda^\prime=\pm} \int \frac{dk_{0} d^3\bm{k}}{\cosh^2\frac{k_{0}}{2T}}  
 \rho(k_{0}, \lambda E_{k})\rho(k_{0}, \lambda^\prime E_{k})\left(
2 \lambda\lambda^{\prime} \frac{k_{z}^2 k_{\perp}^2 }{E_{k}^2} 
 + 2 v_{s}^4  k_{0}^2  \left(1+\lambda \lambda^{\prime} \right) 
 - v_{s}^2 (k_{\perp}^2+2k_{z}^2) k_{0} \frac{\lambda +\lambda^{\prime}}{E_{k}}
\right) .
\end{equation} 
Here we assumed an isotropic speed of sound, $v_{s,\perp}^2 = v_{s,\parallel}^2 = v_s^2$. After averaging over the angular coordinates, the expression simplifies to
\begin{equation}
 \zeta_{\times}^{(B\to 0)} = \frac{1}{64\pi^2 T } \sum_{\lambda,\lambda^\prime=\pm} \int \frac{dk_{0} d^3\bm{k}}{\cosh^2\frac{k_{0}}{2T}}  
 \rho(k_{0}, \lambda E_{k})\rho(k_{0}, \lambda^\prime E_{k})\left(
\frac{4}{15} \lambda\lambda^{\prime} \frac{k^4}{E_{k}^2} 
 + 2 v_{s}^4  k_{0}^2  \left(1+\lambda \lambda^{\prime} \right) 
 - \frac{4}{3} v_{s}^2 k^2 k_{0} \frac{\lambda +\lambda^{\prime}}{E_{k}}
\right) .
\end{equation} 
Using the expressions for the shear and bulk viscosities in Eqs.~(\ref{shear-perp-B0c}) and (\ref{bulk-perp-B0c}), one readily verifies that in the $B \to 0$ limit,
\begin{equation}
\zeta_{\times}^{(B\to 0)} = \zeta_{0} -\frac{2}{3}\eta_{0} .
\label{zeta-cross-0}
\end{equation} 
Thus, although $\zeta_{\times}$ does not vanish as the magnetic field goes to zero, this is not physically significant because it is not an independent transport coefficient in that limit. In a magnetized plasma, however, it becomes a genuinely new and independent transport characteristics.


%

\end{document}